\documentclass[12pt,preprint]{aastex}
\newcommand{\be}{\begin{equation}}
\newcommand{\ee}{\end{equation}}
\newcommand{\bea}{\begin{eqnarray}}
\newcommand{\eea}{\end{eqnarray}}

\newcommand{\ba}{\begin{array}}
\newcommand{\ea}{\end{array}}
\newcommand{\bi}{\begin{itemize}}
\newcommand{\ei}{\end{itemize}}

\newcommand{\ud}[1]{\textrm{d}{#1}}
\newcommand{\der}[2]{\frac{\ud #2}{\ud #1}}

\newcommand{\noi}{\noindent}
\newcommand{\tnm}[1]{\tablenotemark{#1}}
\newcommand{\tnt}[2]{\tablenotetext{#1}{#2}}

\begin{document}

\title{The OPTX Project II: Hard X-ray Luminosity Functions of Active 
Galactic Nuclei for $z\lesssim 5$}
\author{B. Yencho\altaffilmark{1}, A. J. Barger\altaffilmark{2,3,4}, 
L. Trouille\altaffilmark{2}, L. M. Winter\altaffilmark{5}}
\altaffiltext{1}{Department of Physics, University of Wisconsin - Madison, 
1150 University Avenue, Madison, WI 53706}
\altaffiltext{2}{Department of Astronomy, University of Wisconsin - Madison, 
475 N. Charter Street, Madison, WI 53706}
\altaffiltext{3}{Department of Physics and Astronomy, University of Hawaii, 
2505 Correa Road, Honolulu, HI 96822}
\altaffiltext{4}{Institute for Astronomy, University of Hawaii, 
2680 Woodlawn Drive, Honolulu, HI 96822}
\altaffiltext{5}{Department of Astronomy, University of Maryland, 
College Park, MD 20742}

\begin{abstract}
We use the largest, most uniform, and most spectroscopically complete to faint X-ray flux limits {\em Chandra\/} sample to date to construct hard ($2-8$ keV) rest-frame X-ray luminosity functions (HXLFs) of spectroscopically identified active galactic nuclei (AGNs) to $z\sim5$. In addition, we use a new $2-8$~keV local sample selected by the very hard ($14-195$~keV) SWIFT 9-month Burst Alert Telescope (BAT) survey to construct the local $2-8$~keV HXLF.  We do maximum likelihood fits of the combined distant plus local sample (as well as of the distant sample alone) over the redshift intervals $0<z<1.2$, $0<z<3$, and $0<z<5$ using a variety of analytic forms, which we compare with the HXLFs.  We recommend using our luminosity dependent density evolution (LDDE) model fits of the combined distant plus local sample over $0<z<3$ for all the spectroscopically identified sources and for the broad-line AGNs.

\end{abstract}


\section{Introduction}
\label{sec:intro}

Understanding the evolution of the population of active galactic nuclei (AGNs) by means of a luminosity function is an area of continued interest, particularly for determining the accretion history of supermassive black holes (SMBHs) and the relationship to the galaxy population \citep{salucci1999,yu2002,marconi2004,barger2005,merloni2008}.  While AGN surveys have been compiled in a variety of wavelengths, hard X-ray surveys are the least biased by absorption effects and are therefore very well suited to constructing models of the intrinsic space density of AGNs.  In recent years, X-ray luminosity functions have been constructed from both hard \citep{ueda2003,lafranca2005} and soft \citep{miyaji2000,hasinger2005} X-ray samples, while \citet{cowie2003}, \citet{barger2005}, and more recently \citet{silverman2008} have utilized both, taking advantage at redshifts $z\gtrsim 3$ of the depth and small $K$-corrections of soft X-ray data in measuring hard X-ray energies.

\citet{barger2005} fitted their unbinned data over the redshift interval $0<z<1.2$ and found that, in this region, the data were consistent with a pure luminosity evolution (PLE) model, which exhibits the same general shape at each redshift but shows evolution of the characteristic luminosity ``knee'', $L_{*}$.  They noted that PLE does not continue to higher redshifts.  Indeed, it has been shown that a more complex model with both luminosity and density evolution, usually referred to as a luminosity dependent density evolution (LDDE) model, fits the higher redshift data \citep{ueda2003,hasinger2005,lafranca2005,aird2008,ebrero2008,silverman2008}.

Also of interest is the relationship of the broad-line AGNs to the full population.  \citet{steffen2003} and \citet{ueda2003} observed that the fraction of broad-line sources (relative to the full population) increases with X-ray luminosity such that they dominate the AGN population at high luminosities.  This has since been confirmed by a number of groups \citep{barger2005,lafranca2005,akylas2006,tozzi2006,gilli2007,dellaceca2008,hasinger2008} and a similar correlation has been seen with optical luminosity \citep{hao2005,simpson2005}.  Having better HXLFs for both the full and broad-line data, consistent to high redshifts, will aid in understanding this effect.

In this paper we construct full and broad-line $2-8$~keV HXLFs using a data set drawn from the three highly spectroscopically complete X-ray surveys of our OPTX project \citep[see][]{trouille2008}---the \emph{Chandra} Deep Field North (CDF-N), the \emph{Chandra} Large-Area Synoptic X-ray Survey (CLASXS), and the \emph{Chandra} Lockman Area North Survey (CLANS)---as well as from the $\approx1$~Ms \emph{Chandra} Deep Field South (CDF-S) and from an \emph{ASCA} survey.  While \citet{barger2005} previously used the CDF-N, CDF-S, CLASXS, and \emph{ASCA} data to construct their HXLFs, the OPTX project has since improved the spectroscopic completeness of the CDF-N and CLASXS fields and has introduced the CLANS field.  The latter is a series of nine $\sim 70$~ks \emph{Chandra} exposures covering a roughly $\sim 0.6$~deg$^{2}$ area consisting of a total of 761 sources, 533 of which have been spectroscopically observed \citep{trouille2008}.  With all the fields together we have a sample of spectroscopically observed sources which is $\approx 44\%$ larger than that of \citet{barger2005}.  The set of spectroscopically identified sources is similarly enlarged.  With these additional data we can more accurately determine the binned luminosity functions and examine various model fits using maximum likelihood estimates.  We also extrapolate our model fits to compare with a local $2-8$~keV luminosity function that we construct from the first sensitive all-sky very hard ($14-195$~keV) X-ray survey in 28 years, the SWIFT 9-month Burst Alert Telescope (BAT) survey \citep{tueller2008,winter2008,winter2009}.  We then combine the distant sample with the local sample and redo our model fits using the combined data set.  The reason we do the model fits both with and without the local sample is because the BAT survey is not a $2-8$~keV selected sample. However, at the $2-8$~keV fluxes above $\sim 2\times 10^{-11}$~ergs~cm$^{-2}$~s$^{-1}$ the BAT survey should have chosen nearly all of the $2-8$~keV sources over the whole sky, and we can therefore effectively construct a $2-8$~keV sample.  Comparison of the two sets of fits also gives insight into how reliable the extrapolations of the fits are outside the redshift range over which they have been made, since these complex multi-parameter fits may diverge from the observations quite rapidly outside the fitted region.

In \S\ref{sec:data} we briefly describe our distant X-ray sample and the local BAT sample that we use.  In \S\ref{sec:lx} we construct rest-frame luminosities for both samples, and in \S\ref{sec:omega} we determine the effective solid angles.  We construct and analyze our binned luminosity functions in \S\ref{sec:bin} and our unbinned maximum likelihood fits in \S \ref{sec:mlfits}.  We summarize our results in \S\ref{sec:summary}.

In this paper we use $H_{0}=70$~km~s$^{-1}$~Mpc$^{-1}$, $\Omega_{M}=0.3$, and $\Omega_{\Lambda}=0.7$ and the convention that ``log'' is $\log_{10}$.

\section{X-ray Samples}
\label{sec:data}

\subsection{Distant Sample}
\label{secdistant}

We use the $2-8$~keV (hard) and $0.5-2$~keV (soft) fluxes and redshifts of the X-ray samples from 5 highly spectroscopically complete surveys.  These represent a wide range of area and depth regimes, with two very deep but narrow surveys (the CDF-N and CDF-S), one very large but shallow survey (\emph{ASCA}), and two surveys of intermediate depth and coverage (CLASXS and CLANS).  This complementary relationship allows for the construction of a HXLF which best describes the actual population of AGN.

The $\approx$ 2~Ms CDF-N and $\approx$ 1~Ms CDF-S \citep[note that the latter field has recently been deepened by an additional 1~Ms;][]{luo2008} probe the deepest survey volumes over an area of $\sim 0.1$~deg$^{2}$ each.  The CDF-N flux limits are $f_{\rm 2-8~keV}\approx 1.4\times 10^{-16}$ and $f_{\rm 0.5-2~keV}\approx 1.5\times 10^{-17}$~ergs~cm$^{-2}$~s$^{-1}$ \citep{alexander2003}, and we take the CDF-S flux limits to be twice these values \citep{alexander2003,giacconi2002}.  The optical spectra of the X-ray sources in the CDF-N were obtained by \citet{barger2002,barger2003,barger2005} and \citet{trouille2008}, and those of the CDF-S were obtained by \citet{szokoly2004}.  We use the optical spectral classifications given in \citet{trouille2008} and \citet{szokoly2004} to separate out the broad-line AGNs, which are defined therein as all sources which have optical lines with FWHM line widths in excess of $2000$~km~s$^{-1}$.

We include the \emph{ASCA} sample of \citet{akiyama2000} (hereafter referred to as simply ``the \emph{ASCA} sample'') to provide us with higher flux information in the $2-8$~keV band over a large area of $\sim 5$~deg$^{2}$.  The flux limit of this $2-10$~keV sample is $f_{\rm 2-10~keV}\approx 1.0\times 10^{-13}$~ergs~cm$^{-2}$~s$^{-1}$ \citep{akiyama2000}.  We convert the fluxes and flux limit of this sample to the $2-8$~keV band assuming a $\Gamma=1.8$ spectrum.  Of the 30 AGNs in the sample, only 5 are not classified as broad-line AGNs using our above definition.  There is one ambiguous case in which the nominal FWHM of $\sim 1800$~km~s$^{-1}$ would fall below our limit, but this is based only on H$\beta$, so we choose to keep this source as a broad-line AGN.

We also include two intermediate-depth \emph{Chandra} surveys of the Lockman Hole region.  The first is CLASXS, which is an $\sim 0.4$~deg$^2$ field observed and analyzed by our team \citep{yang2004}, and the second is CLANS, which is an $\sim 0.6$~deg$^{2}$ field initially observed by the \emph{Chandra}/SWIRE team (P.I. B.~Wilkes) and subsequently analyzed by our own team \citep{trouille2008}.  For both fields we adopt flux limits of $f_{\rm 2-8~keV}\approx 3\times 10^{-15}$ and $f_{\rm 0.5-2~keV}\approx 5\times 10^{-16}$~ergs~cm$^{-2}$~s$^{-1}$ for our calculations. The optical spectra of the X-ray sources in the CLASXS field were obtained by \citet{steffen2004} and \citet{trouille2008}, and those of the CLANS field were obtained by \citet{trouille2008}.  We again use the optical spectral classifications given in \citet{trouille2008} to separate out the broad-line AGNs. New and updated optical, near-infrared, and mid-infrared data for the CLASXS, CLANS, and CDF-N surveys can be found in \citet{trouille2008}.

In Table~\ref{tab:samples} we give a summary of the number and spectral classes of all the sources in each field.  We note that the \emph{ASCA} field was only observed in the hard band and that there are sources in the other fields which are measured in one band but fall below the detectable level of the other. Thus, in Table~\ref{tab:samples2} we also give the number of sources of each spectral class in each field that have a valid flux (measured and consistent with our chosen flux limits) for the $2-8$~keV and $0.5-2$~keV energy bands separately.  We hereafter refer to these as our hard and soft band samples.

\begin{figure}
\epsscale{0.6}
\plotone{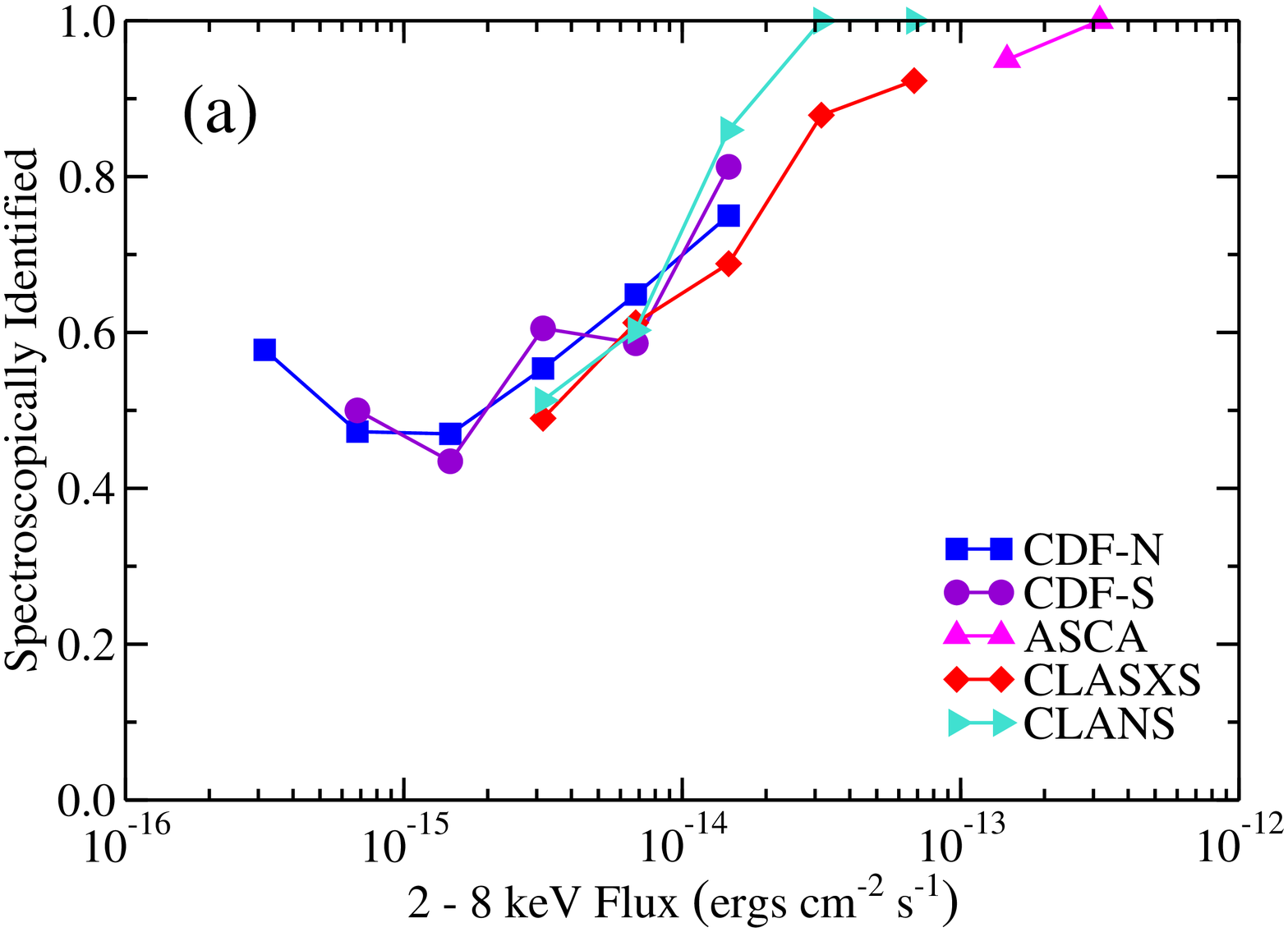} %incompH.eps
\vskip 1.2cm
\plotone{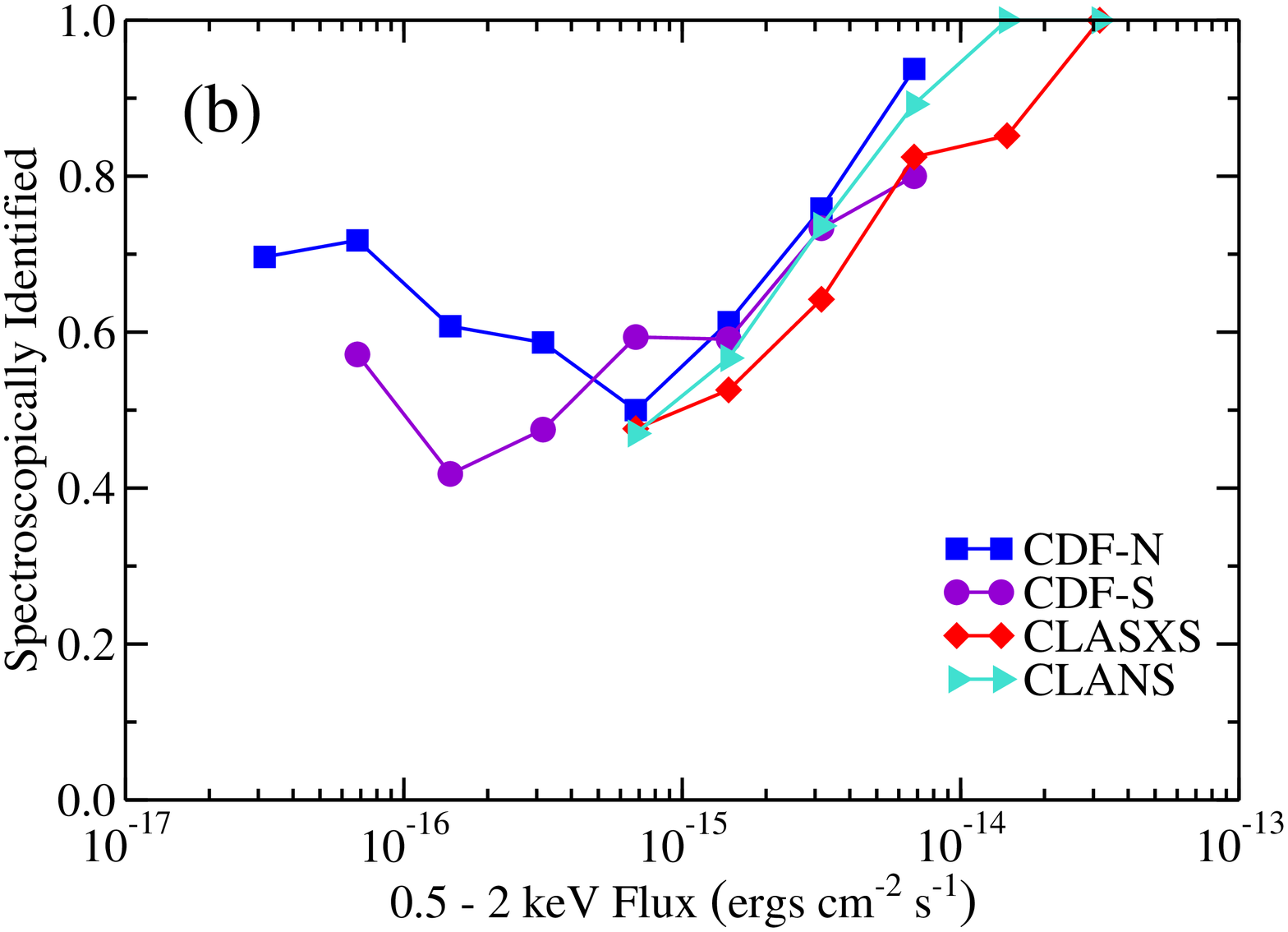} %incompS.eps
\caption{Fraction of observed sources that are spectroscopically identified for each survey in bins of (a) $2-8$~keV flux and (b) $0.5-2$~keV flux.  Only bins which contain 10 or more sources are plotted (with the exception of the highest flux \emph{ASCA} bin, which contains 8 sources but is included due to the survey's limited flux range).  Due to the complicated observing program for the CDF-N, all sources are treated as spectroscopically observed.}
\label{fig:comp}
\end{figure}

The \textit{completeness} of each survey as a function of $2-8$ keV and $0.5-2$ keV flux is given in Figures \ref{fig:comp}a and \ref{fig:comp}b, respectively.  These values were determined by binning the sources according to flux and finding the fraction of the spectroscopically observed sources which are spectroscopically identified.  Due to the complicated observing program of the CDF-N, all of its X-ray sources must be considered as being spectroscopically observed.

While the completeness of each survey is quite good above $10^{-14}$ ergs cm$^{-2}$ s$^{-1}$ in each energy band (70\%-100\%), there is a steady decline with decreasing flux, down to 50\%-60\% in some cases.  This is a result of both the difficulty in obtaining spectra for faint sources and also the intrinsically lower fraction of broad-line AGNs at these fluxes, which are relatively easy to identify.  Perhaps the most difficult sources to identify are the optically normal galaxies at $z \sim 2$ which occupy the intermediate-flux regime.  This dip in completeness is apparent in the two deepest \emph{Chandra} surveys.  The corresponding increase in identifications at the faintest fluxes is associated with our ability to detect low-redshift star forming galaxies.

To avoid problems associated with mixed identification and classification schemes, we choose to use only the spectroscopic identifications, being aware of our incompleteness at lower fluxes, rather than to supplement our spectroscopic identifications with photometric determinations or estimates based on hardness ratios \citep[e.g.,][]{szokoly2004}.  This will manifest itself as an underestimate of the faint end of the luminosity function of the full population, particularly for high-redshift sources, but we will give estimates of the maximum possible error incurred from this approach.

% Number of sources in each survey
\clearpage
\begin{deluxetable}{lccccc}
\tablewidth{0pt}
\tablecaption{X-Ray Samples By Field And Spectral Class}
\startdata
\hline \hline
Category         & CDF-N & CDF-S & \emph{ASCA} & CLASXS & CLANS \\ \hline
Total            & 503   & 346   & 32   & 525    & 761   \\ 
Observed         & 460   & 247   & 32   & 468    & 533   \\
Identified       & 312   & 137   & 31   & 280    & 336   \\
Broad-line       & 39    & 32    & 25   & 103    & 126   \\
High-excitation  & 42    & 23    & 0    & 57     & 92    \\
Star-formers     & 146   & 55    & 5    & 77     & 87    \\
Absorbers        & 71    & 20    & 0    & 23     & 22    \\
Stars            & 14    & 7     & 1    & 20     & 9     \\
\enddata
\label{tab:samples}
\end{deluxetable}

% Numbers w/ usable fluxes
\begin{deluxetable}{lccccc}
\tablewidth{0pt}
\tablecaption{Number of X-Ray Sources With Valid $2-8$~keV/$0.5-2$~keV 
Flux Measurements}
\startdata
\hline \hline
Category         & CDF-N   & CDF-S   & \emph{ASCA} & CLASXS  & CLANS   \\ \hline
Observed         & 299/413 & 185/221 & 31/0 & 304/400 & 378/419 \\
Identified       & 184/288 & 102/122 & 30/0 & 199/253 & 267/283 \\
Broad-line       &  37/38  &  30/32  & 24/0 &  85/103 & 109/123 \\
High-excitation  &  33/39  &  19/21  &  0/0 &  44/48  &  78/74  \\
Star-formers     &  74/134 &  38/48  &  5/0 &  51/63  &  62/59  \\
Absorbers        &  36/63  &  13/14  &  0/0 &  14/19  &  15/18  \\
Stars            &   4/14  &   2/7   &  1/0 &   5/20  &   3/9   \\
\enddata
\label{tab:samples2}
\end{deluxetable}
\clearpage

%(incompH/S.eps)
%
% FIGURE 1
%
\clearpage

\subsection{Local Sample}
\label{seclocal}

For our local comparison sample we start with the 154 sources in the SWIFT 9-month BAT sample of \citet{tueller2008}, for which \citet{winter2008,winter2009} presented the X-ray properties, including the $2-10$~keV fluxes, for 145.  We then exclude the sources within the Galactic plane ($|b|<15^\circ$), where the optical identifications are less complete.  The remaining sky coverage is 30,500~deg$^2$.  \citet{tueller2008} compiled redshifts and optical spectral classifications for the sources.  Here we classify all Seyfert~1s through Seyfert~1.9s as broad-line AGNs. 

\begin{figure}
\epsscale{0.8}
\plotone{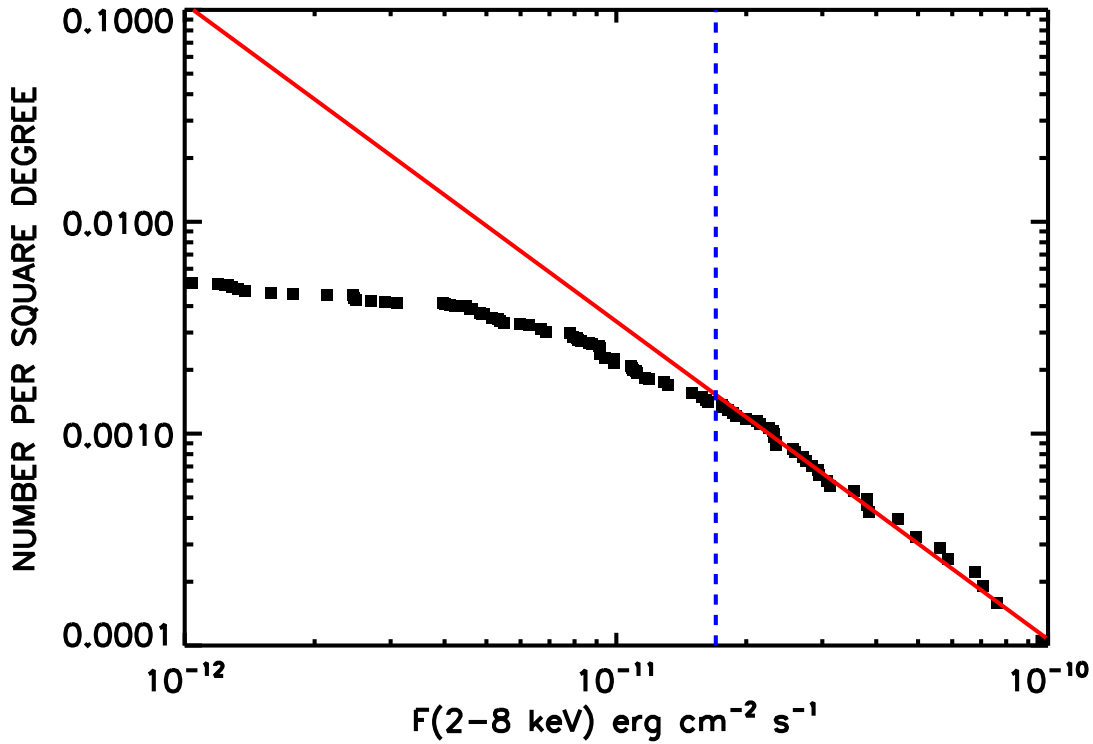}
\caption{Cumulative number counts per square degree vs. $2-8$~keV flux for the $|b|\ge 15^\circ$ BAT sample of Tueller et al.\ (2008; \textit{squares}) using the sky area covered by BAT at the $14-195$~keV flux of each individual source.  The red solid line shows the cumulative distribution of a uniform density of objects (i.e., $-1.5$ slope). The $2-8$~keV sample selected by BAT is complete above about $1.7\times 10^{-11}$~ergs~cm$^{-2}$~s$^{-1}$ (\textit{blue dashed line}).}
\label{fig:ncounts}
\end{figure}

In Figure~\ref{fig:ncounts} we show the cumulative number counts per square degree versus the $2-8$~keV flux (converted from the observed $2-10$~keV flux assuming a photon index $\Gamma=1.8$) for the $|b|\ge 15^\circ$ BAT sample (\textit{squares}).  We used the sky area covered by BAT at the  $14-195$~keV flux of each individual source from \citet{tueller2008}.  The number counts are consistent with the cumulative distribution of a uniform density of objects ($-1.5$ slope; \textit{red solid line}) down to a $2-8$~keV flux of $1.7\times 10^{-11}$~ergs~cm$^{-2}$~s$^{-1}$ (\textit{blue dashed line}), which we take to be the completeness limit of the $2-8$~keV sample selected by BAT.  For our subsequent analysis we only consider the $2-8$~keV sample above this limit, which we hereafter refer to as our local sample.  All 37 of the sources in this sample have redshifts, 35 of which have $z<0.1$.  The median redshift for the sample is $z=0.016$, and the mean redshift is $z=0.023$.  26 of the 37 sources are broad-line AGNs.

%%
%% FIGURE 2 (ncounts.eps) (old f1.eps)
%%\
\clearpage
\clearpage

\section{Rest-Frame Luminosities}
\label{sec:lx}

\subsection{Distant Sample}

To calculate the rest-frame $2-8$~keV luminosities for our spectroscopically identified distant sample (see \S\ref{secdistant} for a description of the sample), a suitable redshift may be chosen such that the luminosities are calculated from the $0.5-2$~keV (rather than the $2-8$~keV band) fluxes for high redshifts.  This is advantageous not only because the \emph{Chandra} images are more sensitive at observed $0.5-2$~keV, but also because at $z = 3$, observed $0.5-2$~keV corresponds to rest-frame $2-8$~keV.

We call the redshift where we change over from using our hard band sample to using our soft band sample $z_{\rm split}$. Assuming a general photon index $\Gamma = 1.8$, we calculate the rest-frame $2-8$~keV luminosities to be

\be
L_{\rm X} =  4 \pi d_{\rm L}^{2}(z)
\times
\left\{
\ba{ll}
F_{\rm hard}(1+z)^{-0.2}
& \textrm{if $z < z_{\rm split}$} \\
F_{\rm soft}\left[\frac{1}{4}(1+z)\right]^{-0.2}
& \textrm{if $z \ge z_{\rm split}$} \\
\ea
\right.
\ee

\noindent where $F_{\rm hard}$ and $F_{\rm soft}$ are the observed $2-8$~keV and $0.5-2$~keV fluxes in the hard and soft band samples, respectively.  For our primary results we choose $z_{\rm split} = 3.0$. The $K$-corrections have different normalizations to characterize the redshift at which observed and rest-frame fluxes are equivalent.

\begin{figure}
\epsscale{0.8}
\plotone{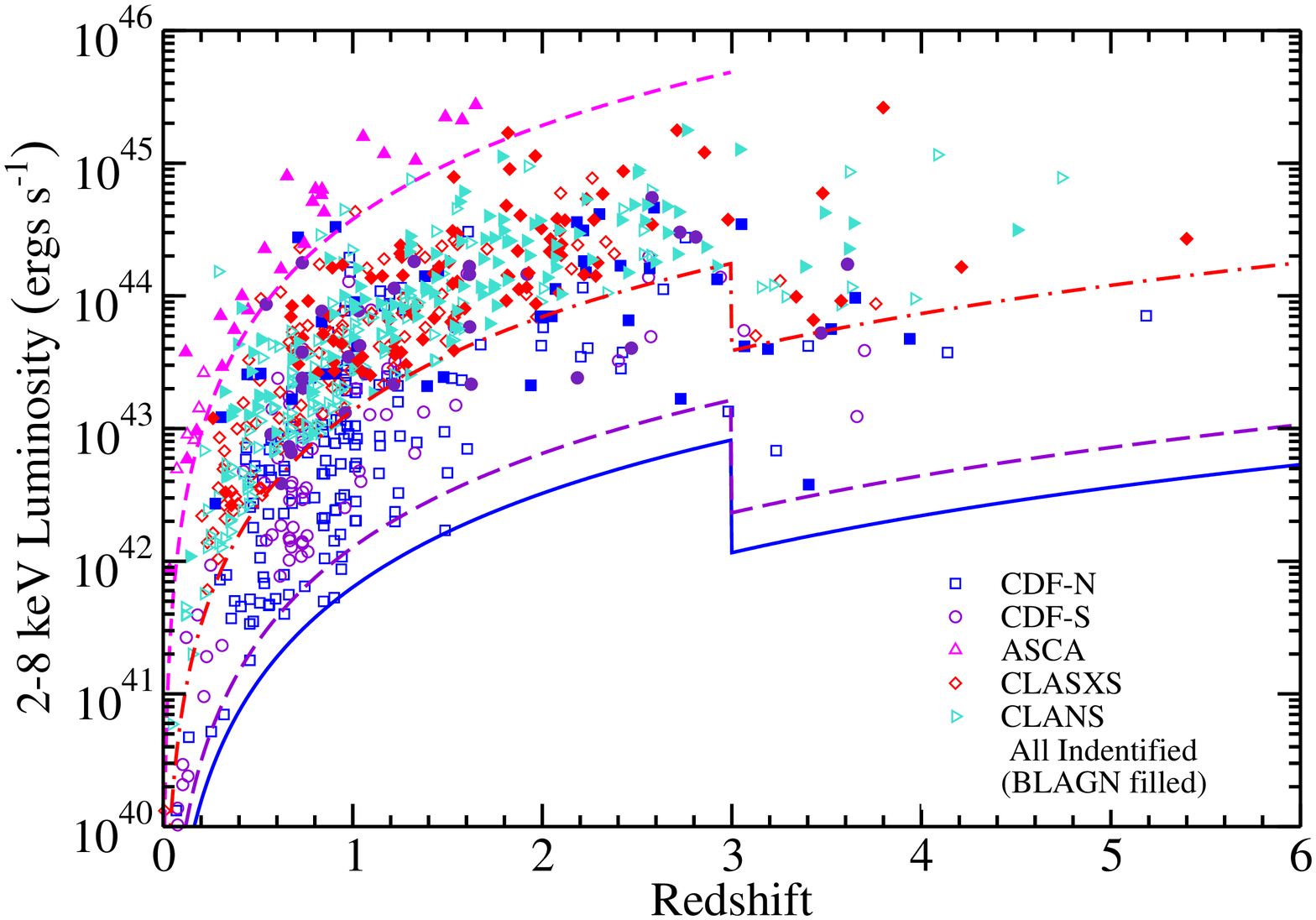}
\caption{Rest-frame $2-8$~keV luminosity vs. redshift for the spectroscopically identified sources in the X-ray surveys listed in the figure legend.  The luminosities were calculated using the $2-8$~keV fluxes from the hard band sample for $z<z_{\rm split}$ and the $0.5-2$~keV fluxes from the soft band sample for $z \ge z_{\rm split}$, with the $K$-corrections determined assuming a general photon index $\Gamma = 1.8$.  The filled symbols designate broad-line AGNs, and the curves correspond to the flux limits of each survey (CDF-N - \textit{blue solid}; CDF-S - \textit{purple long-dashed}; both CLASXS and CLANS - \textit{red dot-dashed}; \emph{ASCA} - \textit{magenta short-dashed}).
}
\label{fig:logLz}
\end{figure}

In Figure \ref{fig:logLz} we show rest-frame $2-8$~keV luminosity versus redshift for the spectroscopically identified sources in the distant sample, labeling each source according to the survey from which it was detected.  The curves correspond to the flux limits for each survey, with the hard band sample flux limits being used for $z<z_{\rm split}$ and the soft band sample flux limits being used for $z \ge z_{\rm split}$.  This reflects our calculation of $L_{\rm X}$ and accounts for the resulting discontinuities.

\begin{figure}
\epsscale{0.6}
\plotone{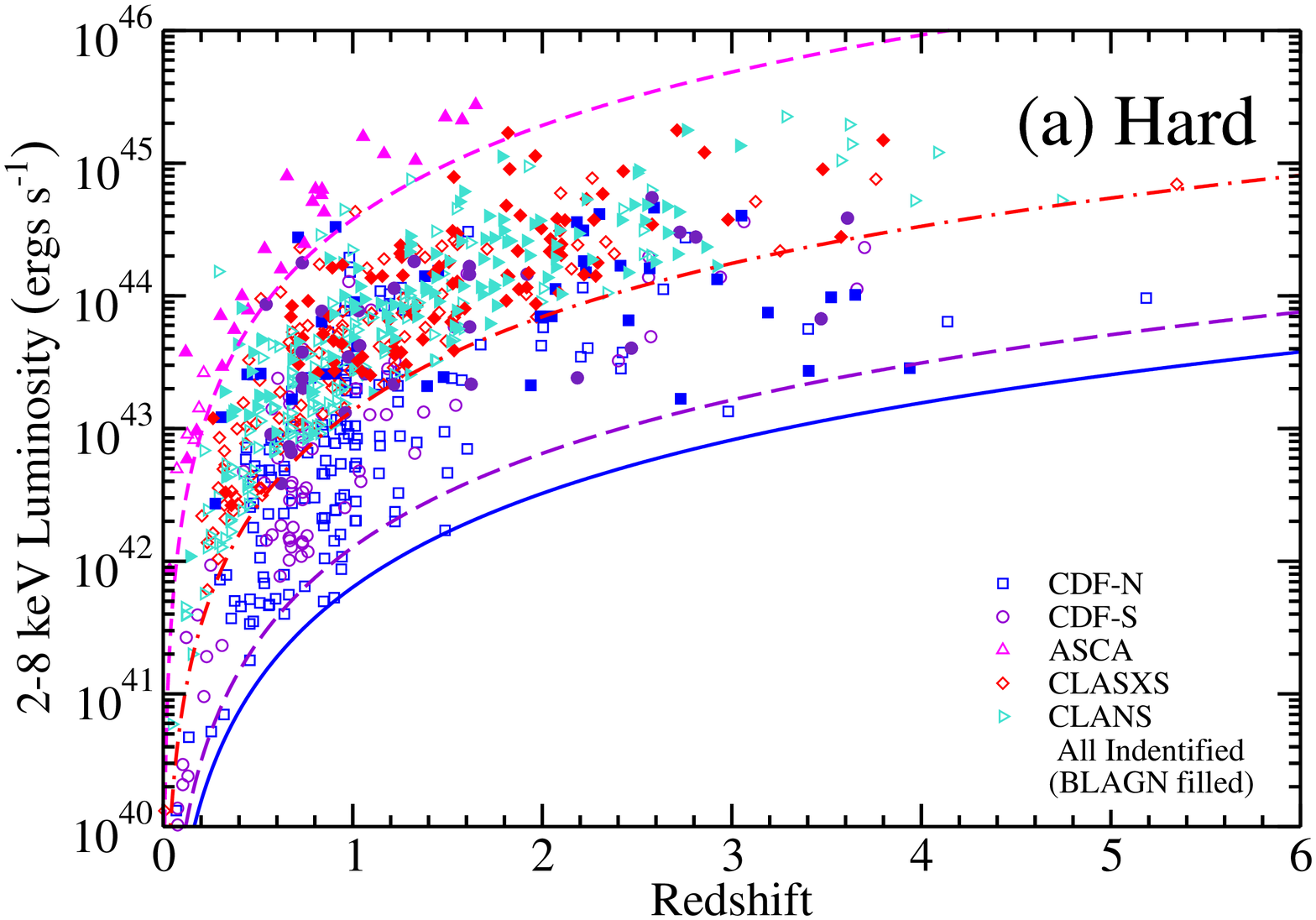} 
\vskip 1.2cm
\plotone{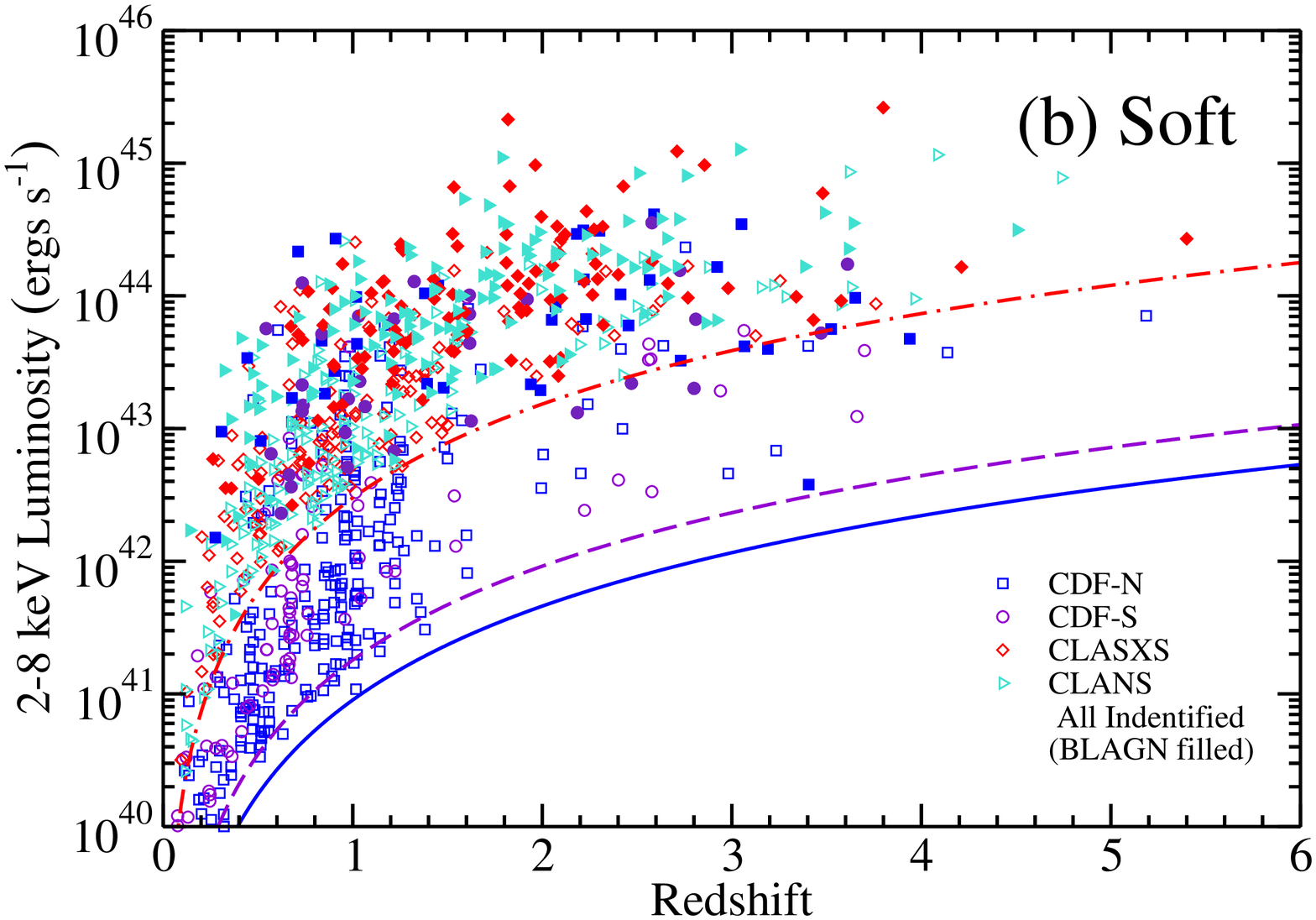} 
\caption{Rest-frame $2-8$~keV luminosity vs. redshift for the spectroscopically identified sources in the X-ray surveys listed in the figure legend.  The luminosities were calculated using (a) the $2-8$~keV fluxes from the hard band sample and (b) the $0.5-2$~keV fluxes from the soft band sample, with the $K$-corrections determined assuming a general photon index $\Gamma = 1.8$.  The filled symbols designate broad-line AGNs, and the curves correspond to the flux limits of each survey (CDF-N - \textit{blue solid}; CDF-S - \textit{purple long-dashed}; both CLASXS and CLANS - \textit{red dot-dashed}; \emph{ASCA} - \textit{magenta short-dashed in panel a}).}
\label{fig:logLz2}
\end{figure}

For comparison, in Figure \ref{fig:logLz2} we show the rest-frame $2-8$~keV luminosities calculated from (a) the hard band and (b) the soft band samples separately, assuming a general photon index $\Gamma=1.8$.  The increased number of high-redshift sources in the soft band sample can be seen, though in both cases the samples are rather sparse at these redshifts.  The soft band sample also has a greater number of low-redshift ($z \lesssim 1$), low-luminosity ($L_{\rm X} \lesssim 10^{42}$ ergs s$^{-1}$) sources relative to the hard band sample, but as we are interested in sources with $L_{\rm X} > 10^{42}$~ergs~s$^{-1}$ \citep[which we assume to be AGNs,][]{zezas1998,moran1999}, we will not be losing too much information by using only the hard band sample at low redshifts.

%%
%% FIGURE 3 (logLz4.eps)  (old f2.eps)
%%
\clearpage

%%
%% FIGURE 4 (logLz4h.eps and logLz4s.eps)  (old f3a/b.eps)
%%
\clearpage

\subsection{Local Sample}

For our local sample (see \S\ref{seclocal} for a description of the sample) we assumed a general photon index $\Gamma=1.8$ and computed the rest-frame $2-8$~keV luminosities from the observed $2-8$~keV fluxes, $F_{\rm hard}$, in this hard band sample using

\be
L_{\rm X} =  4 \pi d_{\rm L}^{2}(z)
F_{\rm hard}(1+z)^{-0.2} \,.
\ee

\begin{figure}
\epsscale{0.8}
\plotone{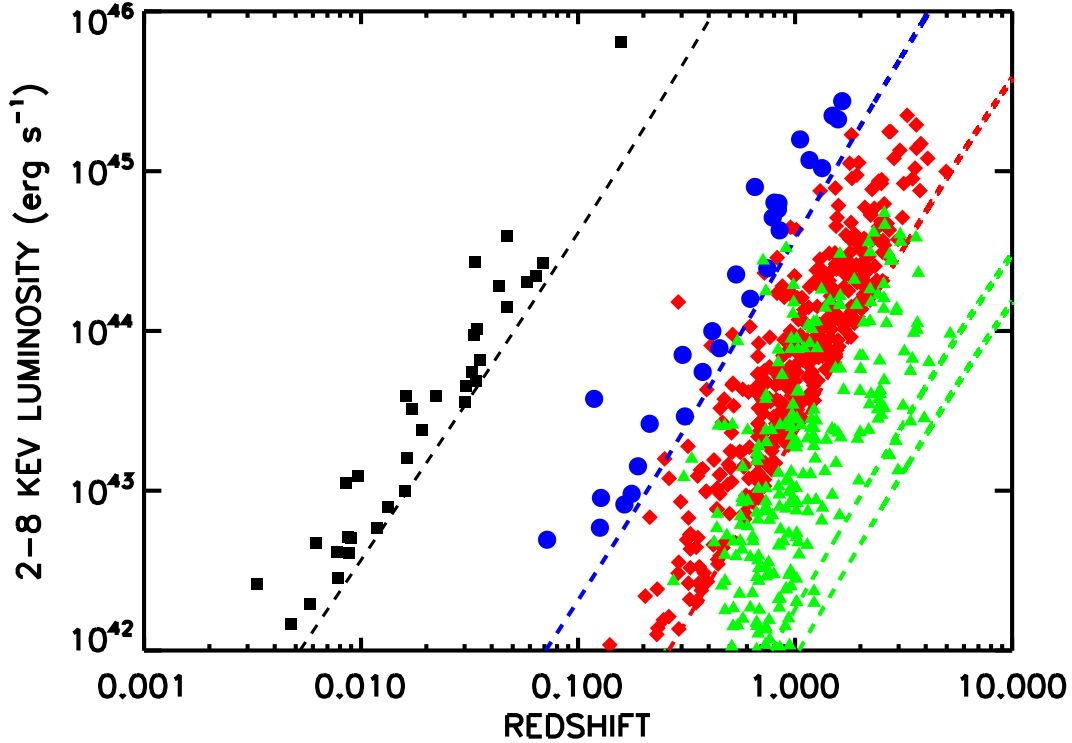}
\caption{Rest-frame $2-8$~keV luminosity vs. redshift for our local sample (\textit{black squares}; note that the blazar at $z=0.859$ is too luminous to appear on the plot) and for the spectroscopically identified sources in the \emph{ASCA} (\textit{blue circles}), CLASXS plus CLANS (\textit{red diamonds}), and CDF-N plus CDF-S (\textit{green triangles}) hard band samples.  The luminosities were calculated using the $2-8$~keV fluxes, with the $K$-corrections determined assuming a general photon index $\Gamma = 1.8$.  The colored dashed lines show the luminosity limits corresponding to the flux limits of the surveys (from left to right: local sample; \emph{ASCA}; CLASXS/CLANS; CDF-S; CDF-N).}
\label{fig:lxz}
\end{figure}

\noindent In Figure~\ref{fig:lxz} we show the rest-frame $2-8$~keV luminosities versus redshift for our local sample (\textit{black squares}) and for the spectroscopically identified sources in the \emph{ASCA} (\textit{blue circles}), CLASXS plus CLANS (\textit{red diamonds}), and CDF-N plus CDF-S (\textit{green triangles}) hard band samples.  We show the luminosity limits corresponding to the flux limits of the surveys as the colored dashed lines.  Our local sample is almost entirely a $z\le 0.1$ sample (with two blazar exceptions, one of which is too luminous to appear on the plot) and is therefore essentially disjoint and complementary to our distant sample. The distant surveys are also complementary to one another and provide continuous coverage in both luminosity and redshift over the $z<5$ redshift range where we are measuring the HXLFs.

%%
%% FIGURE 5 (old f4.eps)
\clearpage%%
\clearpage

\section{Effective Solid Angles}
\label{sec:omega}

\subsection{Distant Sample}
The sensitivity of X-ray detectors typically varies with the off-axis angle of the central pointing, resulting in an effective solid angle which is smaller for fainter sources.  Our approach to estimating the effective solid angle (as a function of flux) of our total hard and soft distant samples is as follows.  We first bin our hard and soft samples separately according to (log) $2-8$~keV and $0.5-2$~keV flux, respectively.  In each bin, we then compare the number of objects which are spectroscopically observed to independent determinations of expected number per unit area per unit flux\footnote{Note that for all of the fields except the CDF-N, the spectroscopic observations were targeted at the X-ray sources.  Thus, we can assume that the unobserved sources would be like the observed sources and exclude them.  In the method we are using to determine the area, when we remove the sources that were not observed, the area decreases accordingly.  For the CDF-N, the observing strategy was more complex, and we need to treat all of the sources in that field as if they were observed.}.  These \textit{differential number counts}, here represented as $n(F)$, use simple modeling of detector sensitivity to correct for the described bias.  We therefore estimate the effective solid angle for a bin with central value $\log F_{c} \equiv 0.5 (\log F_{1}+\log F_{2})$ and observed count $N$ as

\be
\Omega^{\rm bin}_{\rm X_{i}}(F_{c}) \approx \frac{N}{
n_{\rm X_{i}}(F_{\rm c}) \, \times
(F_{2}-F_{1}) 
}
\ee

\noi where $\rm X_{i}$ =  $2-8$ keV or $0.5-2$ keV and the differential number counts $n_{\rm X_{i}}$ (in units of deg$^{-2}$ per $10^{-15}$~ergs~cm$^{-2}$~s$^{-1}$) are given by

\bea
n_{\rm H}(F_{\rm H}) &=& \left\{
\begin{array}{ll}
(39 \pm 5)(F_{\rm H}/F_{b})^{-2.57 \pm 0.22} & \qquad \qquad
\textrm{if $F_{\rm H} > F_{b}$} \\ \nonumber
(32 \pm 2)(F_{\rm H}/F_{b})^{-1.63 \pm 0.05} & \qquad \qquad
\textrm{if $F_{\rm H} < F_{b}$} \\
\end{array}
\right.
%\ee
%\be
\\
n_{\rm S}(F_{\rm S}) &=& \left\{
\begin{array}{ll}
(12.49 \pm 0.02)(F_{\rm S}/F_{b})^{-2.5} & \qquad
\textrm{if $F_{\rm S} > F_{b}$} \\
(12.49 \pm 0.02)(F_{\rm S}/F_{b})^{-1.7 \pm 0.02} & \qquad
\textrm{if $F_{\rm S} < F_{b}$} \\
\end{array}
\right.
\eea

\noi for the $H=2-8$~keV  \citep{cowie2002} and $S=0.5-2$~keV  bands \citep{yang2004}, respectively\footnote{\citet{yang2004} found it difficult to precisely fit the slope of their $0.5-2$~keV data for  fluxes $>10^{-14}$~ergs~cm$^{-2}$~s$^{-1}$ but found it to be consistent with a fixed value of $\alpha=2.5$, resulting in our lack of errors on this value.}.  In each case, the break flux occurs at $F_{b} = 10^{-14}$~ergs~cm$^{-2}$~s$^{-1}$.

\begin{figure}
\epsscale{0.6}
\plotone{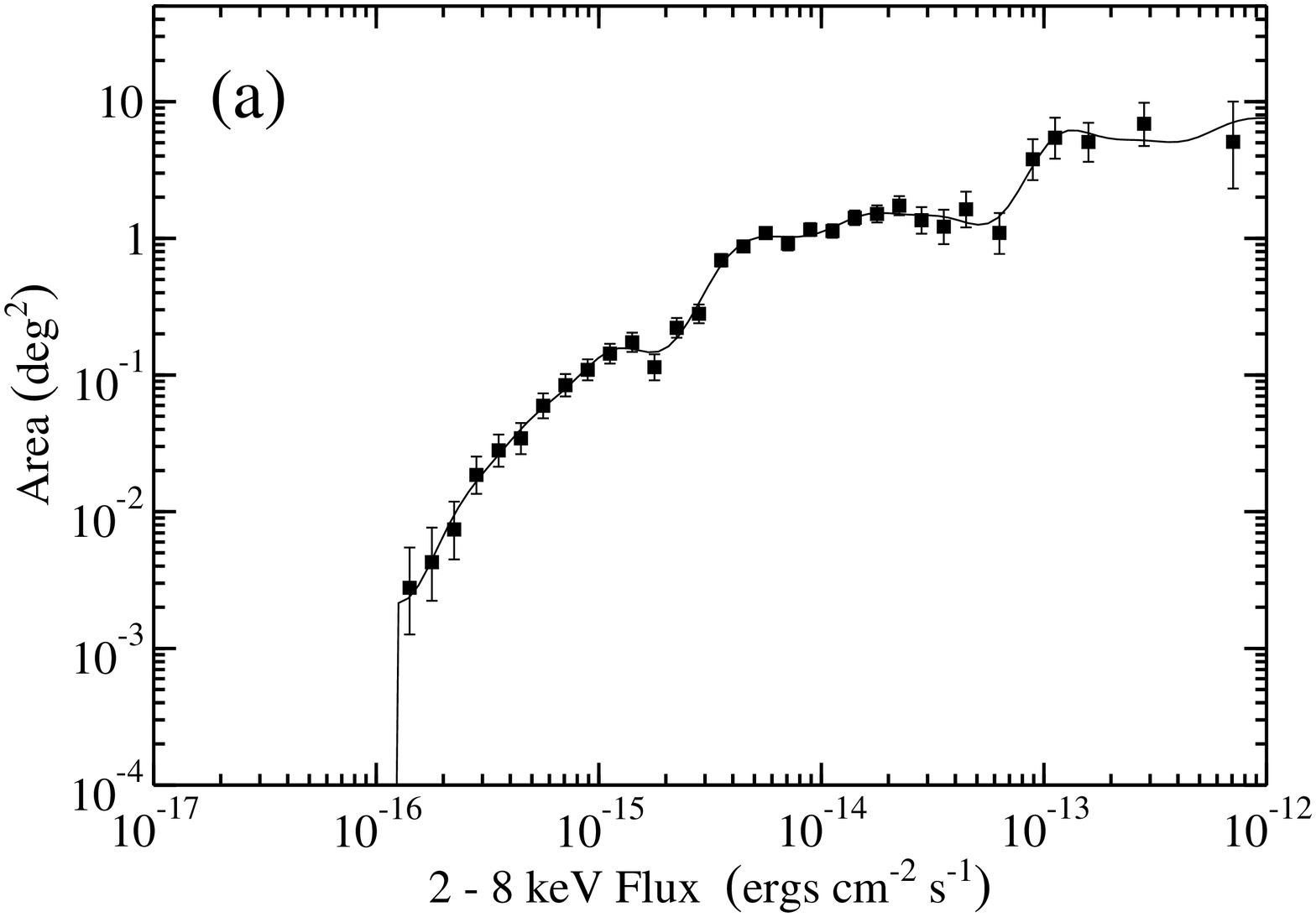} 
\vskip 1.2cm
\plotone{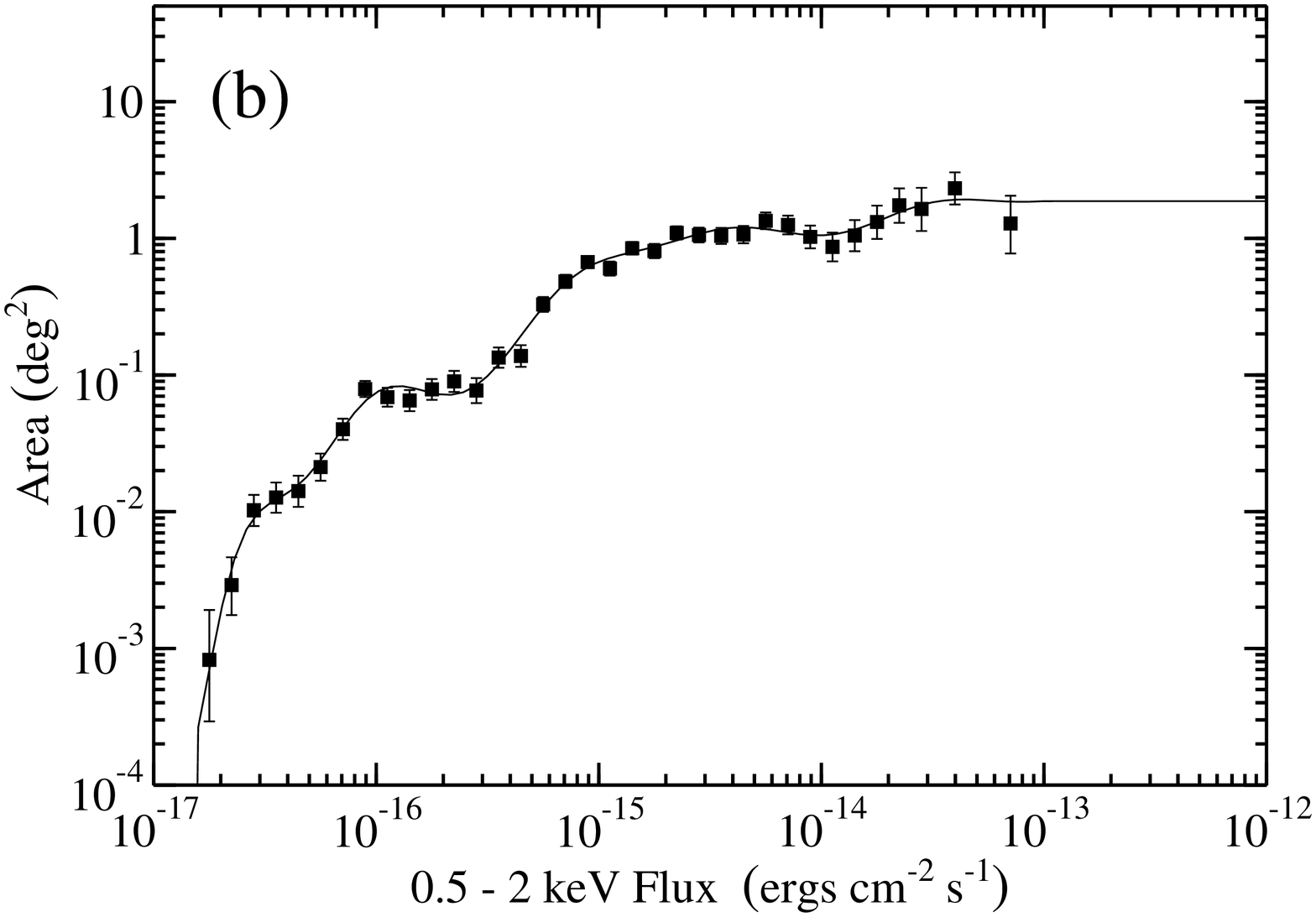} 
\caption{The total effective solid angle of the distant surveys as a function of (a) $2-8$~keV and (b) $0.5-2$~keV flux.  The points represent the binned data, with the error bars indicating the $1~\sigma$ Poissonian uncertainties. The solid curves are harmonic series fits to the data.}
\label{fig:effarea}
\end{figure}

The resulting binned estimates are given in Figure~\ref{fig:effarea}, where the $1 \sigma$ Poissonian uncertainties are based on the number of objects in each bin.  To reduce the dependence on binning, particularly for flux bins at the extremes of the distribution where the data are scarce, it is advantageous to fit the binned data with simple curves that show little variation with the details of the binning procedure.  Smooth curves also facilitate the numerical computations necessary for estimating the luminosity function.  Therefore we use the binned data to fit a truncated harmonic series in $\log \Omega - \log F$ space according to a general linear least squares approach.  We only assume these fits to be valid over the range of fluxes covered by the data.  For fluxes above the brightest flux bin, we assume the effective solid angle is approximately constant and equal to the high flux endpoint of the fit; for fluxes lower than the faintest flux limit of our samples which are thus observationally inaccesible, we assign a value of zero to the curves.  This process is performed separately for the hard and soft samples and the resulting curves are presented in  Figure~\ref{fig:effarea}.  Calling these curves $\Omega_{\rm H}$ and $\Omega_{\rm S}$, we then define a single function which contains information from both distributions:

\be
\Omega(\log L_{\rm X},z) = \left\{
\ba{ll}
\Omega_{\rm H}(F_{\rm H})
& \textrm{if $z < z_{\rm split}$} \\
\Omega_{\rm S}(F_{\rm S})
& \textrm{if $z \ge z_{\rm split}$} \\
\ea
\right.
\label{fig:omega}
\ee

\noi Here $F_{\rm H}$ and $F_{\rm S}$ are the hard and soft fluxes that would be observed for a source with a rest-frame $2-8$~keV luminosity $L_{\rm X}$ at redshift $z$, assuming a general photon index $\Gamma = 1.8$. This use of $z_{\rm split}$ to define a single effective solid angle describing both the hard and soft samples follows from our definition of $L_{\rm X}$.

We also investigated how the use of different authors' fits to their differential number counts data would affect our results.  Using the $n_{\rm H}(F_{\rm H})$ and $n_{\rm S}(F_{\rm S})$ derived from the $2-8$ keV and $0.5-2$ keV samples of both \citet{trouille2008} and \citet{kim2007}, we recalculated $\Omega_{\rm \rm H}(F_{\rm H})$ and $\Omega_{\rm S}(F_{\rm S})$ according to the above method.  We find these forms of $\Omega_{\rm H}(F_{\rm H})$ to be in very good agreement with our previous determination, but see some small discrepancies at both the faint and bright ends of $\Omega_{\rm S}(F_{\rm S})$.  However, these differences had no significant effect on our final results, given the uncertainties.

\subsection{Local Sample}

Since the local sample is selected in the BAT high-energy band, determining the solid angle versus $2-8$~keV flux relation is not straightforward.  Above our chosen flux limit of $1.7\times10^{-11}$~ergs~cm$^{-2}$~s$^{-1}$ in the $2-8$~keV band the situation is simpler, since most of the sources could be found across the full area of 30,500~deg$^2$ for the sky at $|b|\ge 15^\circ$.  However, even at these $2-8$~keV fluxes, about a third of the sources with $z=0-0.1$ (those with higher flux ratios) have covered areas which are smaller than this, with the smallest being 14,400~deg$^2$. We have therefore tried two area versus $2-8$~keV flux relations. In the first we assumed a constant area of 27,000~deg$^2$ equal to the mean area in the sample. In the second, we derived an area versus $2-8$~keV relation from the area versus $14-195$~keV relation of \citet{tueller2008} by a simple conversion of the $14-195$~keV flux to the corresponding $2-8$~keV flux found when assuming a fixed ratio of the two fluxes equal to the median ratio of 0.53.  These fixed and flux dependent $2-8$~keV area relations give almost identical results for the luminosity functions, and in the following we show the results for the fixed area case only.

%%
%% FIGURE 6  (old f5a/b.eps)
%%
\clearpage%%
\clearpage

\section{Hard X-ray Luminosity Functions: Binned}
\label{sec:bin}

\subsection{\citet{page2000} Method}

Let the hard X-ray luminosity function, defined as the number of objects per comoving volume per (log) luminosity, be given by

\be
\phi(\log L_{\rm X},z) = \der{\log L_{\rm X}}{\Phi(\log L_{\rm X},z)} \,.
\ee 

\noi We wish to estimate the value of this function in a bin of data defined by $\log L_{1} < \log L < \log L_{2}$ and $z_{1} < z < z_{2}$. We begin by following the derivation of the binned luminosity function by \citet{page2000}.  Given $M$ different surveys, each with a unique flux limit and effective survey area, the expected number of objects in this bin would be given by 

\be
\langle N \rangle = 
\sum_{j}^{M} \left[
\int_{\log L_{1}}^{\log L_{2}}
\int_{z_{1}}^{z_{\rm max}^{j}(\log L)}
\phi(\log L,z) \, \Omega_{j}(\log L,z) \, \der{z}{V}(z)
\, \ud{z} \, \ud{\log L}
\right] \,,
\ee

\noi where $\ud{V}/\ud{z}$ is the differential volume corresponding to a solid angle of 1~steradian, $\Omega_{j}(\log L,z) = \Omega_{j}(F)$ is the effective survey area of the $j^{\rm th}$ survey, and $z_{\rm max}^{j}(\log L)$ is the maximum possible redshift within the bin for which an object of luminosity $L$ could be detected by survey $j$ (given the restrictions of that survey's flux limit). If we now assume that $\phi$ varies little over this bin, we can factor it out of the integral, yielding

\be
\phi \approx \frac{N}{
\sum_{j} 
\left[
\int_{\log L_{1}}^{\log L_{2}}
\int_{z_{1}}^{z_{\rm max}^{j}(\log L)}
\Omega_{j}(\log L,z) 
(\ud{V}/\ud{z}) \, \ud{z} \, \ud{\log L}
\right]
} \,,
\label{eq:coherent}
\ee

\noi where $N$ is the observed number of objects in the bin.  Written in this 
form, the estimate is an implementation of the method of \citet{page2000} which adheres to the ``coherent'' addition of samples described by \citet{avni1980}.  We can manipulate this further, though, to find a form more suitable for our data.

Consider that each $\Omega_{j}(F)$ is a binned function derived according to the procedure given in \S \ref{sec:omega} but using only the data of the $j^{\rm th}$ survey.  Then, by construction, each of these is equal to zero for any flux fainter than its flux limit:

\be
\Omega_{j}\left( F<F_{\rm limit}^{j} \right) = 0 \,.
\ee

\noi This means that the $z_{\rm max}^{j}(\log L)$ provide redundant information, and we may set the upper limit of the redshift integrals to $z_{2}$ without approximation. The sum can then be moved inside the integral:

\be
\phi \approx \frac{N}{
\int_{\log L_{1}}^{\log L_{2}}
\int_{z_{1}}^{z_{2}}
\left[
\sum_{j} \Omega_{j}(\log L,z) 
\right]
(\ud{V}/\ud{z}) \, \ud{z} \, \ud{\log L}
} \,.
\ee

\noi The factor in brackets can be identified with the total effective survey area from Equation \ref{fig:omega}. The final form, then, for our estimate is 

\be
\phi \approx \frac{N}{
\int_{\log L_{1}}^{\log L_{2}}
\int_{z_{1}}^{z_{2}}
\Omega(\log L,z) 
%(\ud{V}/\ud{z})
\der{z}{V}(z)
\, \ud{z} \, \ud{\log L}
} \,,
\label{eq:phi}
\ee

\noindent with the 1~$\sigma$ Poissonian uncertainty \citep{gehrels1986} on this value determined by the number of objects in the bin.  This form is convenient because it packages all the information concerning the limitations of the surveys into a single quantity, $\Omega(\log L,z)$, rather than in the limits of multiple integrals.  Also, given the binning required to determine the effective, flux-dependent solid angle for a particular survey, a direct calculation of the total solid angle for all surveys allows for the best possible statistics in each bin.  This is the primary advantage of Equation~\ref{eq:phi} over Equation~\ref{eq:coherent}.

\subsection{Results for the Distant Sample}
\label{sec:results-bin}

\begin{figure}
\epsscale{0.9}
\centering
\plottwo{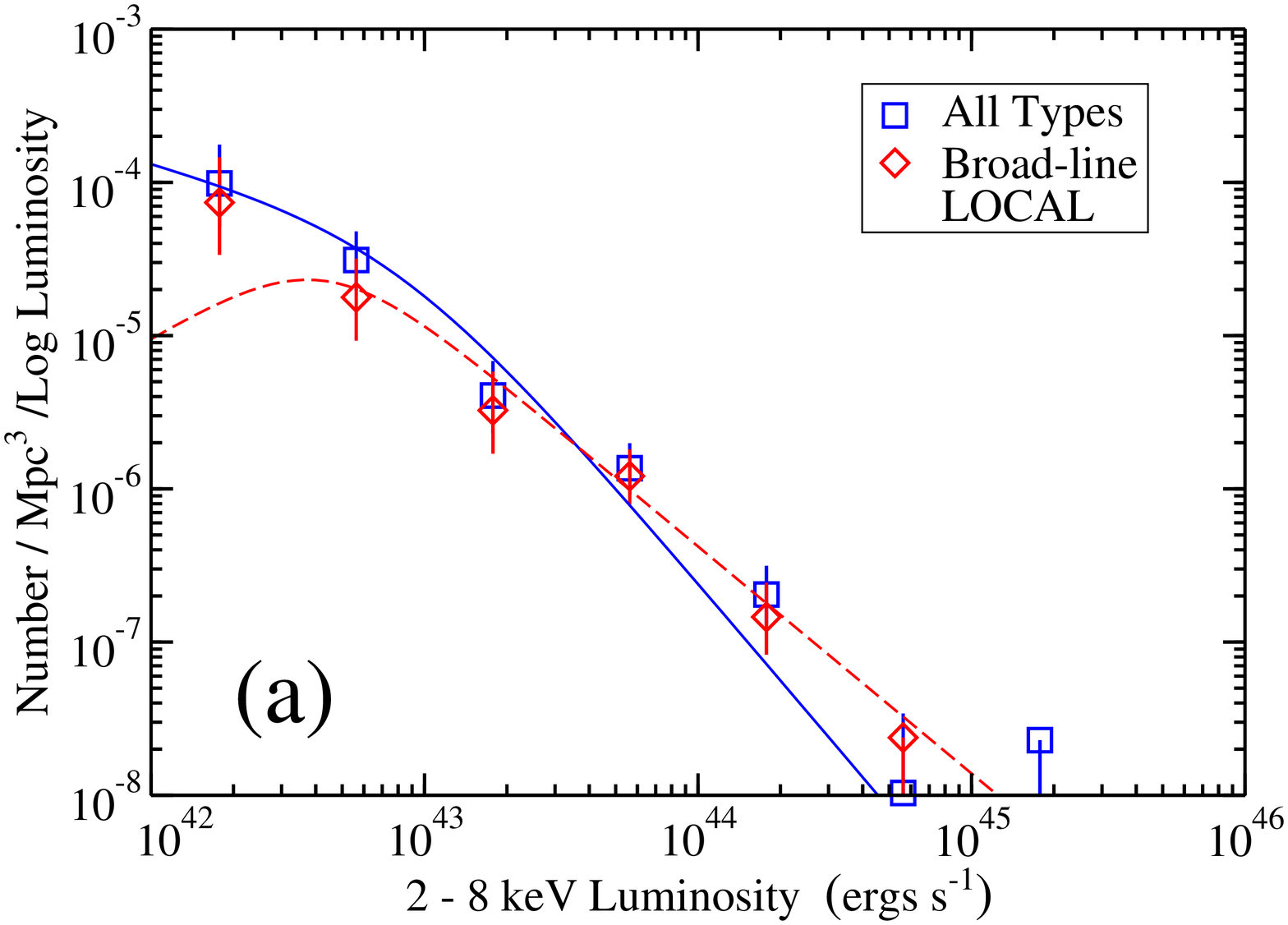}{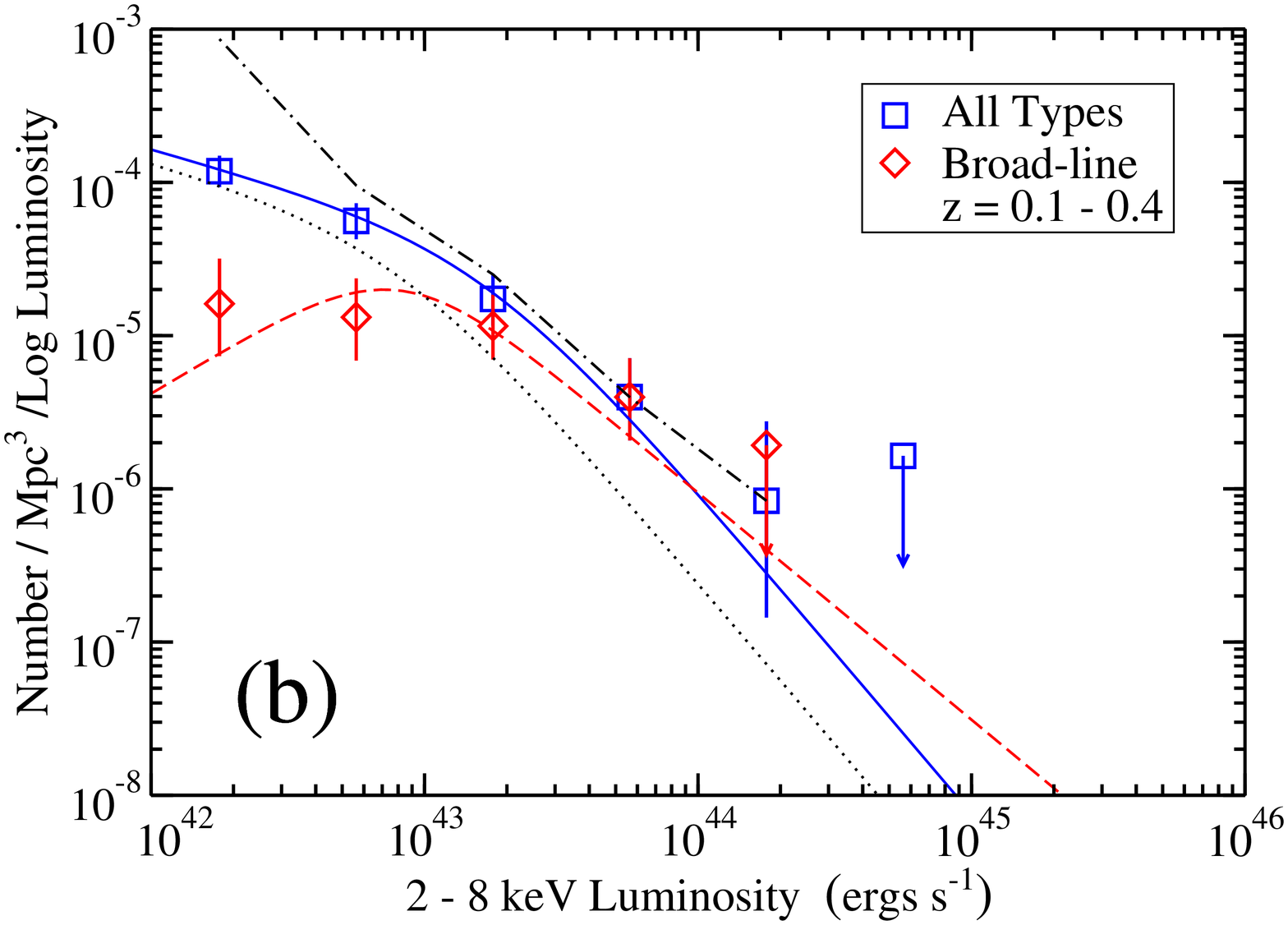}
\vskip 0.8cm
\plottwo{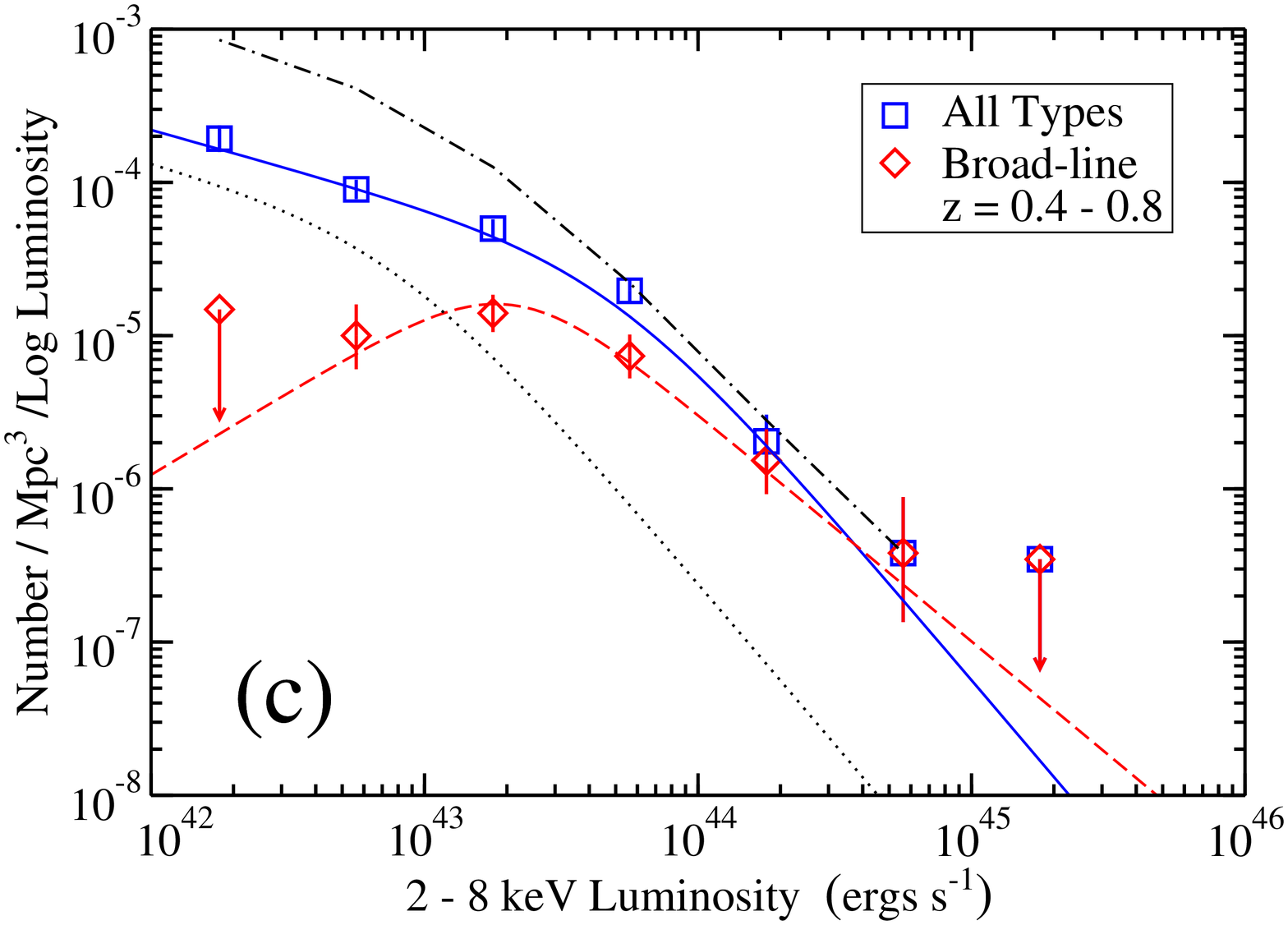}{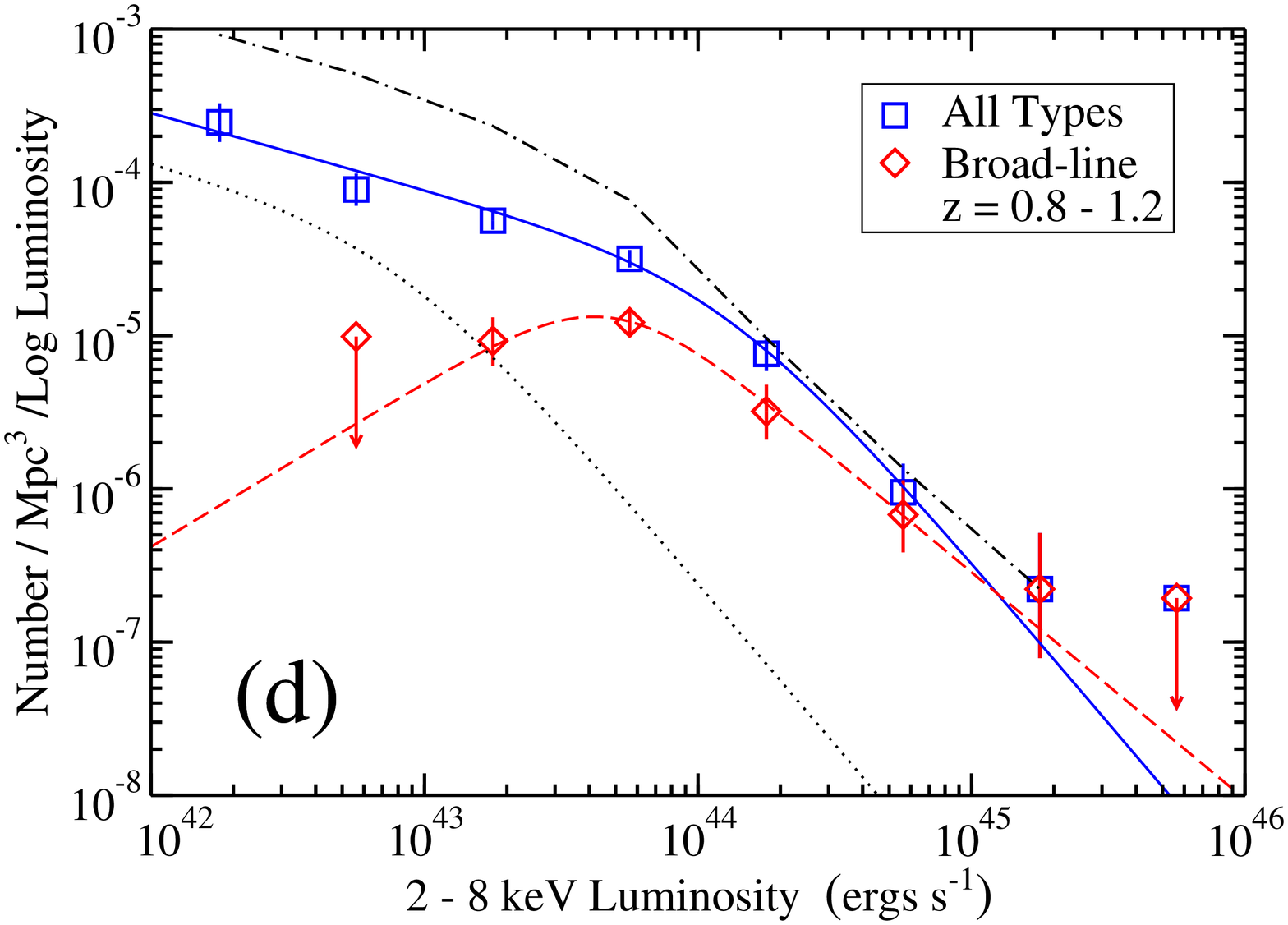}
\vskip 0.8cm
\plottwo{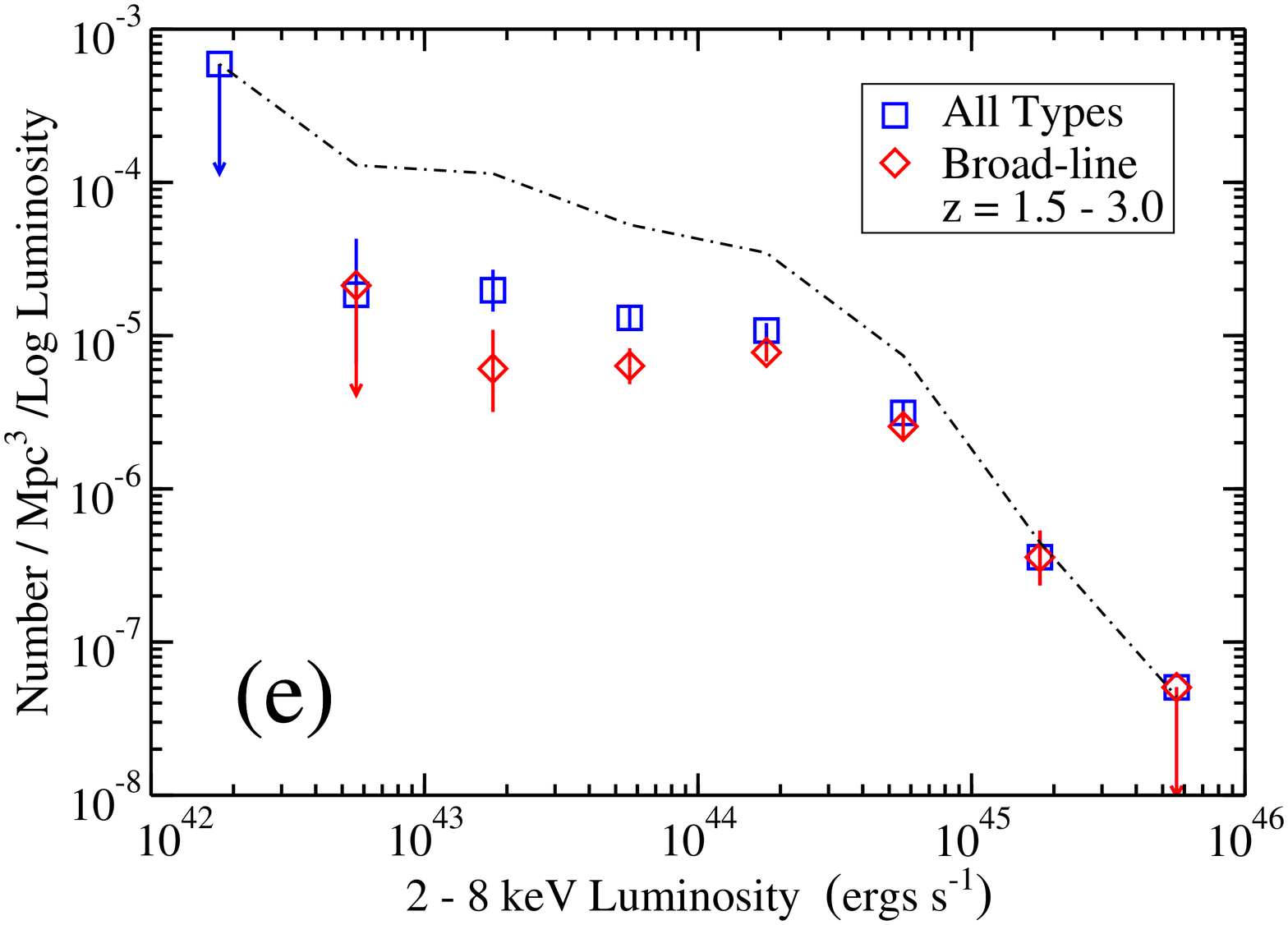}{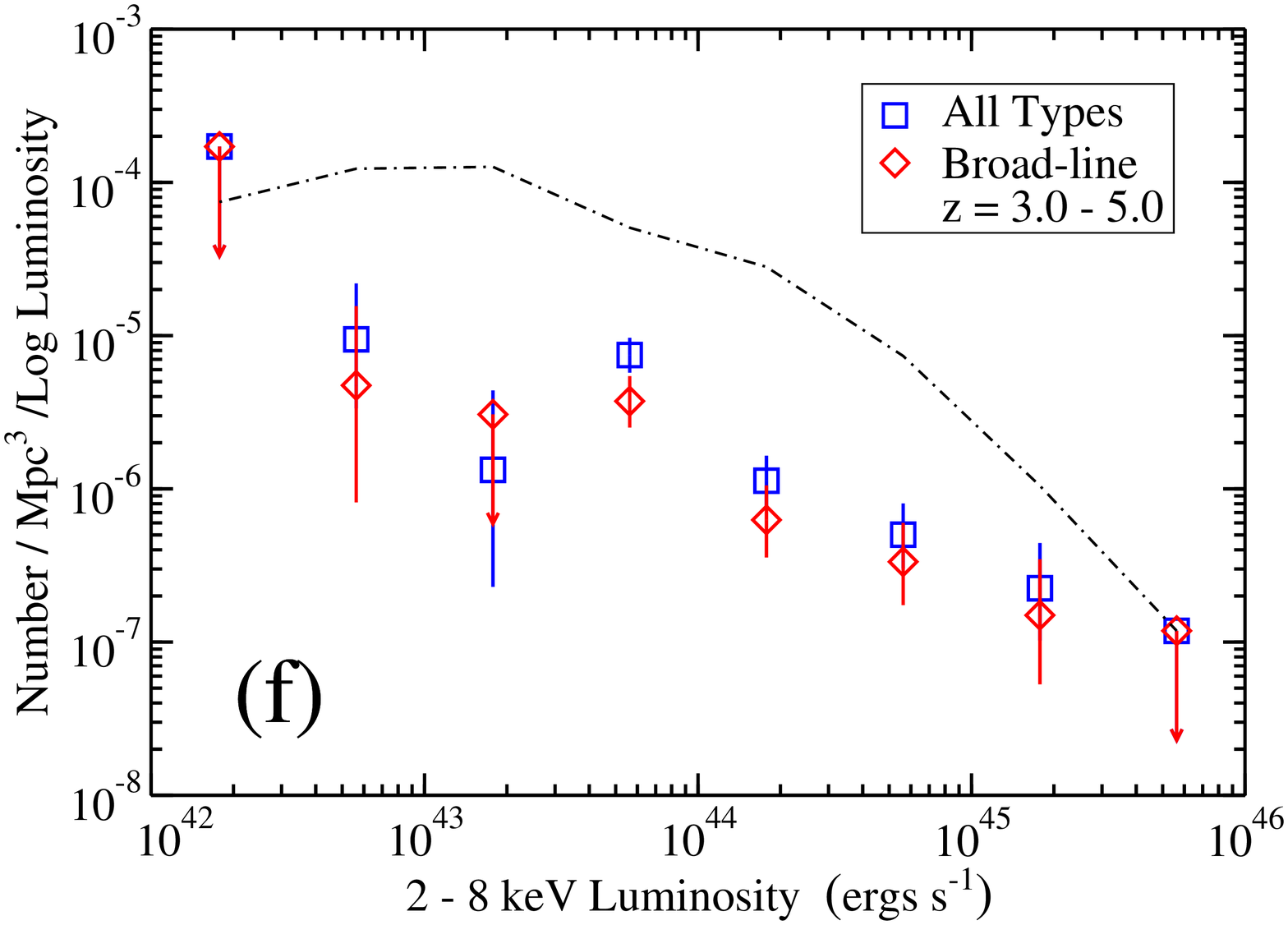}
\caption{Binned rest-frame $2-8$ keV luminosity functions for our (a) $0<z<0.1$ local sample (see \S\ref{sec:localhxlf}) and for our distant sample in the low-redshift bins (b) $0.1 < z < 0.4$, (c) $0.4 < z < 0.8$, and (d) $0.8 < z < 1.2$ and in the high-redshift bins (e) $1.5 < z < 3.0$ and (f) $3.0<z<5.0$.  Blue squares (red diamonds) denote the HXLFs for all spectroscopically identified sources (broad-line AGNs) in each redshift bin.  Error bars indicate the 1~$\sigma$ Poissonian uncertainties based on the number of objects in each bin, while arrows denote $90\%$ (2.3 object) upper limits. The black dot-dashed curves in each panel (except a) give the maximal HXLFs found by assigning the central redshift of that redshift bin to all the spectroscopically observed but unidentified sources.  The blue solid and red dashed (b$-$d) curves, respectively, show the maximum likelihood fits over $10^{42}< L_{\rm X}<10^{46}$~ergs~s$^{-1}$ and $0<z<1.2$ using the \textbf{ILDE model} (plotted at the geometric mean of each redshift bin) for the full and broad-line data of the distant sample (see \S\ref{sec:results-fits}).  The $z=0$ extrapolation of the fits is shown in (a), and the extrapolation of the full sample fit is also included in each higher redshift panel (except e and f) as a black dotted curve.}
\label{fig:lfA}
\end{figure}

We calculate binned HXLFs according to the above procedure for the distant sample having $L_{\rm X}>10^{42}$~ergs~s$^{-1}$ in three low-redshift bins and two high-redshift bins.  In each redshift bin we calculate HXLFs both for all spectroscopically identified sources together (\textit{blue squares}; hereafter we will sometimes refer to this sample as either ``full'' or ``all spectral types'') and for broad-line AGNs alone (\textit{red diamonds}).  These binned HXLFs can be seen in Figures~\ref{fig:lfA}b$-$f.  They are consistent with \citet{barger2005}.  Moreover, as in that paper, the differences between the full and broad-line HXLFs are clear:  At high X-ray luminosities the full and broad-line HXLFs are consistent, but while the full HXLF continues to rise with lower X-ray luminosities, the broad-line HXLF appears to fall off.

% ILDE
%%
%% FIGURE 7  (old f6a-f.eps)
%%
\clearpage%%
\clearpage

We also calculate maximal luminosity functions in each redshift bin to estimate the effect of incompleteness on our results.  For a particular bin, this is achieved by assigning all the spectroscopically observed but unidentified sources with valid flux measurements a redshift at the center of the bin.  This calculation is then performed for each redshift bin, giving results which are therefore incompatible but which provide an estimate of the largest possible error due to incompleteness.  These results can be seen as the black dot-dashed curves in Figures~\ref{fig:lfA}b$-$f.

Incompleteness will have some effect in each redshift bin but is most likely to bias the results for higher redshifts due to the difficulty in getting spectroscopic identifications of low luminosity, high-redshift sources.  Large incompleteness at these redshifts would help explain the flattening of the HXLFs relative to the low-redshift bins.  \citet{barger2005} assigned photometric redshifts (where possible) to the unidentified sources in the CDF-N and CDF-S (working from \citealp{zheng2004} for the latter) and computed a ``spectroscopic plus photometric'' luminosity function.  They found the largest discrepancy with their spectroscopic-only results to be in the $1.5<z<3.0$ bin, which suggests that most of the unidentified sources in those samples lie in this redshift bin.  Indeed, a look at Figure~\ref{fig:logLz2} admits the possibility of an underdensity of low luminosity sources in this redshift region due to the incompleteness of these samples.

Another way to understand how our incompleteness could affect our results is to impose stricter flux limits such that the remaining sample is $\sim 90\%$ spectroscopically complete.  We performed such a cut to each survey separately and then recalculated our full and broad-line AGN HXLFs.  There was little variation in either, within the uncertainties, at low redshifts.  At higher redshifts, the bright end showed no significant change, but the self-imposed limits resulted in a complete lack of faint end information.  This affirms our conjecture that our incompleteness is related to faint, high-redshift sources, but it does not provide us with an estimate of the magnitude of the effect.

While incompleteness is a major source of uncertainty when considering the full HXLF at high redshifts, we do not believe the broad-line AGN data suffer from such a problem.  This is because broad-line AGNs are easier to detect and measure spectroscopically.  We therefore believe our broad-line AGN data to be fairly complete at least out to $z=3$, where we have enough data to calculate a reasonably constrained luminosity function.

An important characteristic of our HXLF for the broad-line AGNs is the significant turnover in number density below the luminosity break.  A similar turnover in number density is not seen the soft X-ray luminosity function of ``type-1 AGNs'' of \citet{hasinger2005}.  In contrast to our use of pure spectroscopic redshifts and optical classifications, \citet{hasinger2005}  use both spectroscopic and photometric redshifts, and they adopt a mixed classification scheme for ``type-1 AGNs'' that includes any object optically classified as a broad-line AGN as well as any object with $L_{\rm X}>10^{42}$ ergs s$^{-1}$ and a \emph{Chandra} - specific hardness ratio $(C_{\rm hard}-C_{\rm soft})/(C_{\rm hard}+C_{\rm soft})<-0.2$, where $C_{\rm hard}$ and $C_{\rm soft}$ are the $2-8$~keV and $0.5-2$~keV count rates \citep[see][]{szokoly2004}.  (They also employ a flux-dependent completeness correction in the calculation of their HXLF, which is known to be biased when the remaining sources have not been identified for non-random reasons, though these corrections are small in their case.)  They adopt this mixed classification scheme because according to the simple unified AGN model \citep[e.g.,][]{antonucci1993} there should be a one-to-one correspondence between the optically unobscured (broad-line) and the X-ray unobscured (type-1) AGNs, with a similar relation for the obscured sources.  However, studies have shown that there is a mismatch between optical and X-ray identifications of $10-20\%$ (e.g., \citealp{garcet2007} and references therein; Trouille et al.~2009, in preparation).  The physical mechanism for this mismatch is not yet fully understood, and thus we avoid the use of mixed classifications, which may introduce unknown biases and effects.

It has been suggested that pure optical classification schemes may misidentify broad-line AGNs as optically ``normal'' galaxies (i.e., our star former and absorber classes), particularly at lower luminosities \citep{gilli2007}.  For example, it has been shown that local Seyfert 2 galaxies may not be properly identified as such due to the host galaxy light overwhelming that of the nuclear region \citep{moran2002,cardamone2007}.  \citet{barger2005} tested this galaxy dilution hypothesis using their CDF-N data by comparing the nuclear UV magnitudes of the sources with the $0.5-2$ keV fluxes, which are known to be strongly correlated for broad-line AGNs \citep[e.g.,][]{zamorani1981}.  They found that the optically identified broad-line AGNs showed this correlation, while the other classes did not (their nuclei were much weaker relative to their X-ray light than would be expected if they were similar to the broad-line AGNs).  Recently, \citet{cowie2009} combined Galaxy Evolution Explorer \citep[GALEX;][]{martin2005} observations with the CLASXS, CDF-N, and CDF-S X-ray samples to determine the ionizing flux from $z \sim 1$ AGNs, and they found that only the broad-line AGNs are ionizers; all of the non--broad-line AGNs are UV faint.  From these two lines of evidence we conclude that we are not misidentifying sources as non--broad-line AGNs when they are really broad-line AGNs, as one might have expected to happen if the broad lines were not visible spectroscopically due to dilution by the host galaxy.

We also considered the effect that choosing a different $z_{\rm split}$, the redshift separating our use of the hard band and soft band samples, would have on our results.  Instead of the $z_{\rm split}=3.0$ we have been using, we tried a value of $z_{\rm split}=1.5$.  The consequence of this is that both the $1.5<z<3.0$ and $3<z<5$ redshift bins of our HXLF are determined by our soft band sample, rather than just the latter.  We find that this does not have a significant effect on the results of this or other sections, and thus we claim our analysis to be independent of the choice of $z_{\rm split}$.

%%
%% FIGURE 8  (old f8.eps)
%%
\begin{figure}
\epsscale{0.8}
\plotone{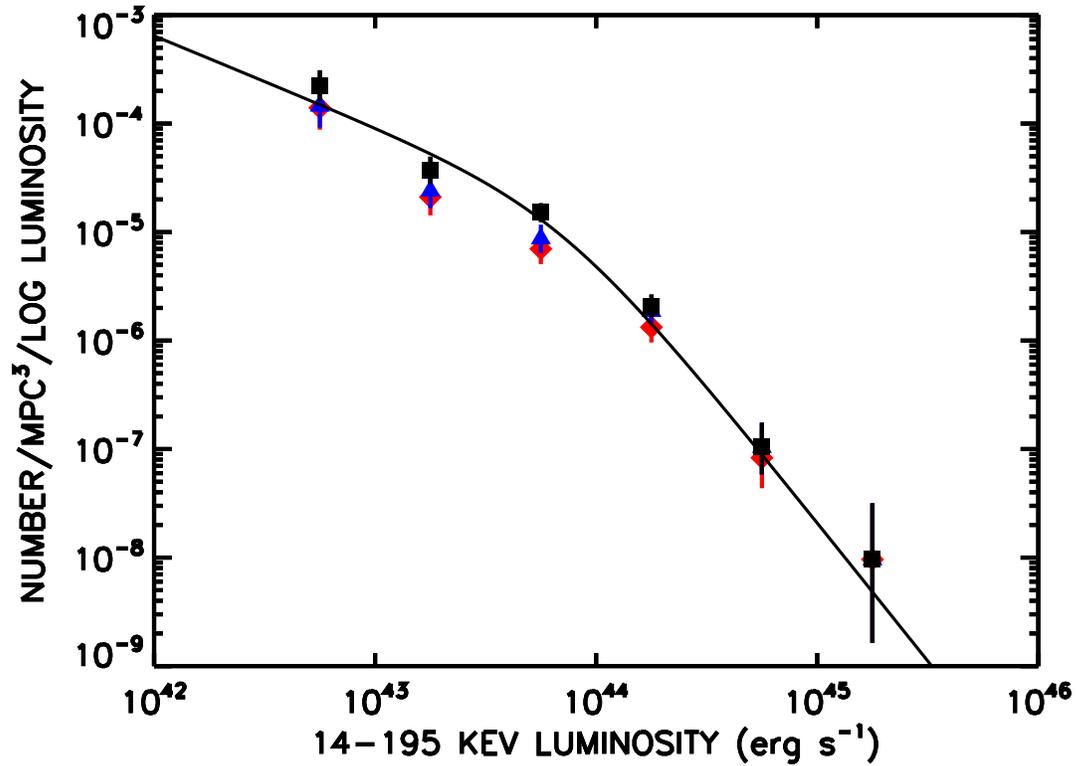}
\caption{Local binned $14-195$~keV luminosity function 
(\textit{black squares}) and \citet{tueller2008}'s analytic approximation (\textit{black curve}) to this quantity for the $|b|\ge 15^\circ$ BAT sample.  The red diamonds show the binned $14-195$~keV luminosity function for the broad-line AGNs, and the blue triangles show the same for the X-ray soft sources (defined as having a $2-8$~keV flux to $14-195$~keV of 0.14).}
\label{fig:localvHXLF}
\end{figure}
\clearpage

\subsection{Results for the Local Sample}
\label{sec:localhxlf}

Using the $|b|\ge 15^\circ$ BAT sample, we constructed binned $14-195$~keV luminosity functions for the full sample, for the broad-line AGNs, and for the X-ray soft sources (which we take to be  sources for which the ratio of the $2-8$~keV flux to the $14-195$~keV flux is greater than 0.14).  In Figure~\ref{fig:localvHXLF} we compare the luminosity functions for the broad-line AGNs (\textit{red diamonds}) and for the X-ray soft sources (\textit{blue triangles}) with the total luminosity function (\textit{black squares}), as well as with the analytic approximation of \citet{tueller2008} (see their Eq.~1 and Table~2; \textit{black curve}). Our result for the total luminosity function reproduces that of \citet{tueller2008}.

We see that the two selections, optical spectral class and X-ray softness, give very similar results. Nearly all of the most luminous sources are X-ray soft broad-line AGNs, and it is only below the break in the luminosity function at a  luminosity of $10^{44}$~ergs~s$^{-1}$ ($14-195$~keV) that there are significant numbers of X-ray hard sources. Even here the bulk of the sources are X-ray soft broad-line AGNs, which suggests that there are not huge numbers of highly absorbed sources.  As \citet{tueller2008} pointed out, the total luminosity function has a steeper faint-end slope than the extrapolation of the $2-8$~keV luminosity function to $z=0$ from \citet{barger2005}; however, the faint-end slope of the X-ray soft sample is flatter and hence in better agreement with \citet{barger2005}.  This is reassuring, since it is the X-ray soft sources which will appear in the $2-8$~keV samples.

\begin{figure}
\epsscale{0.8}
\plotone{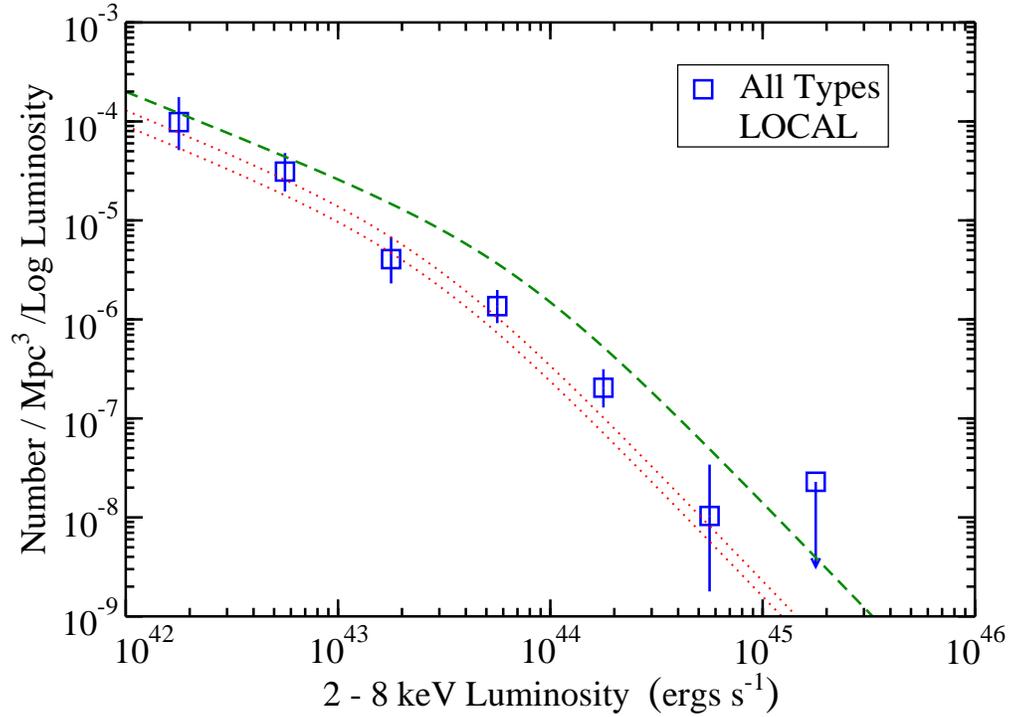}
\caption{Binned rest-frame $2-8$ keV luminosity function for our $0<z<0.1$ local sample (see \S\ref{sec:localhxlf}) (\textit{blue squares}).  The red dotted curves show the local $3-20$~keV luminosity function of \citet{sazonovrev2004} using \textit{RXTE} data (with and without their incompleteness correction of 1.4) after converting to $2-8$~keV assuming a general photon index $\Gamma=1.8$, while the green dashed curve is the $z=0$ extrapolation of the LDDE model fit of \citet{ueda2003}, converted from $2-10$~keV to $2-8$~keV luminosity assuming $\Gamma=1.8$.}
\label{fig:localHXLF}
\end{figure}

We can directly quantify this by constructing the $2-8$~keV luminosity function from the BAT selected local $2-8$~keV sample (see \S\ref{seclocal} for a description; this is the sample we have been referring to as our local sample).  We show this luminosity function in Figure~\ref{fig:lfA}a for all $L_{\rm X}>10^{42}$~ergs~s$^{-1}$ spectroscopically identified sources (\textit{blue squares}) and broad-line AGNs (\textit{red diamonds}) in the redshift bin $0<z<0.1$.  The comparison of the binned data with the extrapolations of our maximum likelihood fits will be discussed later. In Figure~\ref{fig:localHXLF} we compare the luminosity function for the full sample (\textit{blue squares}) with the local $3-20$~keV luminosity function of \citet{sazonovrev2004} from \textit{RXTE} data (with and without their incompleteness correction of 1.4; \textit{red dotted curves}) after conversion to $2-8$~keV assuming a general photon index $\Gamma=1.8$.  We find good agreement.

In the construction of their HXLF, \citet{ueda2003} include a sample of 49 AGNs from two \textit{HEAO 1} surveys which cover a region of $L_{\rm X}-z$ space very similar to that of our local BAT sample.  Our chosen flux limit of $1.7\times10^{-11}$~ergs~cm$^{-2}$~s$^{-1}$ in the $2-8$~keV band is comparable to that of the \textit{HEAO 1} MC-LASS survey and only slightly deeper than the $\sim 2.2 \times 10^{-11}$~ergs~cm$^{-2}$~s$^{-1}$ limit (when converted to $2-8$~keV) of the \textit{HEAO 1} A-2 survey, from which 28 of their 49 sources are drawn.  These samples are included in the \citet{ueda2003} total evolutionary HXLF model fits, which are fit out to $z=3$ using a total sample of 247 AGNs.  We show the $z=0$ extrapolation of their LDDE model fit in Figure \ref{fig:localHXLF} (\textit{green dashed curve}).  The curve agrees with our local HXLF only at the faintest luminosities and otherwise lies above our values and the curves of \citet{sazonovrev2004}.  This may in part be due to the fact that \citet{ueda2003} fit for an ``intrinsic'' HXLF before absorption effects, but it is also likely related to their limited sample size and to problems associated with getting an accurate representation of the local luminosity function while simultaneously fitting over an extended redshift region.

%%
%% FIGURE 9  (old f9.eps)
%%
\clearpage%%
\clearpage

\section{Hard X-ray Luminosity Functions: Maximum Likelihood Fits}
\label{sec:mlfits}

It is often of particular use to find an analytic form for a HXLF independent of the binning procedure described above.  A standard approach is to assume some model with unknown parameter dependence and use the maximum likelihood method to estimate the values which best fit the data \citep{marshall1983}.  To do this, we define the function

\be
N(\log L_{\rm X},z) = 
\phi_{\rm model}(\log L_{\rm X},z) \,
\Omega(\log L_{\rm X},z) \,
\der{z}{V}(z) \,,
\ee

\noi where $\phi_{\rm model}$ is our parameter dependent model for the HXLF.  $N(\log L_{\rm X},z)$ is simply the number density of AGNs per logarithmic luminosity per redshift predicted by the model.  The best-fit parameter values in a region of the HXLF defined by $\log L_{\rm min} < \log L_{\rm X} < \log L_{\rm max}$ and $z_{\rm min} < z < z_{\rm max}$ are those which minimize the function

\be
\mathcal{L} = -2 \sum_{i} \ln \left[
\frac{N(\log L_{{\rm X}, i},z_{i})}{
\int \int
N(\log L_{\rm X},z) \, \ud{\log L_{\rm X} \, \ud{z}}
}
\right] \,,
\ee

\noi whose sum is over all objects in the sample which occupy this region of $(\log L_{\rm X},z)$ space \citep[e.g.,][]{miyaji2000}.  The 1~$\sigma$ errors for the parameters are then estimated as the increase/decrease needed to achieve  $\Delta \mathcal{L} = 1$ from the minimized value, $\mathcal{L}_{\rm min}$, while simultaneously minimizing the function over all other parameters.

The function $\mathcal{L}$ will be independent of any overall normalization of $\phi_{\rm model}$.  To estimate such a parameter, we require that the number of objects predicted by the model in the given region be equal to the observed number.  The 1~$\sigma$ error will then be given by Poisson statistics, with a value $\approx A/\sqrt{N}$ for normalization $A$ and total number of objects $N$.

Unlike a $\chi^{2}$-fitting, the maximum likelihood method does not provide any estimate of goodness-of-fit.  Therefore, we use Kolmogorov-Smirnov (K-S) tests as independent estimates of the quality of the fits \citep{press1992}.  Beginning with the distribution defined by $N(\log L_{\rm X},z)$, 1-D K-S tests are performed on both the $\log L_{\rm X}$ and $z$ distributions (found by integrating out one of the components and normalizing) and 2-D tests are performed on the full, normalized distribution.  It should be noted that these K-S tests are designed to give the probability that a sample distribution may have been drawn from a given model distribution.  This interpretation is not strictly true when any of the model parameters have been estimated from the sample distribution, as is the case here, and therefore these tests should only be thought of as a rough estimate of the goodness-of-fit.

\subsection{Models}
\label{sec:models}

For each of the models we consider, the $z=0$ behavior is represented as a double power law \citep{picc1982}

\be
\der{\log L_{\rm X}}{\Phi(L_{\rm X},z=0)}=
A\left[
\left(\frac{L_{\rm X}}{L_{*}}\right)^{\gamma_{1}} + 
\left(\frac{L_{\rm X}}{L_{*}}\right)^{\gamma_{2}}
\right]^{-1} \,.
\ee

\noi The manner in which this form evolves with redshift characterizes each class of models.  The two most basic involve an evolution of either $L_{*}$ or $A$.  These models are referred to as pure luminosity evolution (PLE) and pure density evolution (PDE) models, respectively, and have been studied in the X-ray samples by various authors \citep{miyaji2000,ueda2003,hasinger2005,barger2005}.  We begin by considering the form used by \citet{barger2005}\footnote{We actually utilize a slight redefinition of the parameters used in \citet{barger2005}, who fit for the values of the luminosity ``knee'', $L_{0}$, and the density normalization, $a_{0}$, at $z=1$.  We define equivalent $z=0$ parameters, $L_{*}$ and $A$, such that $L_{*}=L_{0}(1/2)^{pL}$ and $A=a_{0}(1/2)^{pD}$.  This has no effect on the actual behavior of the model.}, which allows for independent evolution of each:

\be
\der{\log L_{\rm X}}{\Phi(L_{\rm X},z)}=
\der{\log L_{\rm X}}{\Phi(L_{\rm X}/e_{L}(z),0)}\times e_{D}(z) \,,
\ee

\noi where

\be
e_{L}(z) = (1+z)^{pL} \,,
\ee
\be
e_{D}(z) = (1+z)^{pD} \,.
\ee

\noi This model, which for the purpose of this paper we will refer to as the independent luminosity and density evolution (ILDE) model, has six free parameters, four of which determine the $z=0$ behavior ($L_{*},A,\gamma_{1},\gamma_{2}$) and two of which contribute to the redshift evolution of the luminosity knee and overall density normalization ($pL,pD$).

We also consider a luminosity dependent density evolution (LDDE) model with the general evolutionary behavior given by

\be
\der{\log L_{\rm X}}{\Phi(L_{\rm X},z)}=
\der{\log L_{\rm X}}{\Phi(L_{\rm X},0)}\times e(z,L_{\rm X}) \,.
\ee

\noi A useful form of $e(z,L_{\rm X})$ introduced by \citet{ueda2003} is

\be
e(z,L_{\rm X}) = \left\{
\ba{ll}
(1+z)^{p1}                     & z \le z_{c}(L_{\rm X}) \\
e(z_{c})[(1+z)/(1+z_{c})]^{p2} & z   > z_{c}(L_{\rm X}) \\
\ea
\right.
\,,
\ee

\be
z_{c}(L_{\rm X}) = \left\{
\ba{ll}
z_{c}^{*} (L_{\rm X}/L_{a})^{\alpha} & L_{\rm X} \le L_{a} \\
z_{c}^{*}                            & L_{\rm X}   > L_{a} \\
\ea
\right.
\,.
\ee

\noi  This model has nine parameters: the four $z=0$ parameters and five others governing the redshift evolution ($p1,p2,z_{c}^{*},L_{a},\alpha$).  It is usually the case that $p1>0$ and $p2<0$, allowing for a density increasing as 
$(1+z)^{p1}$ up to some luminosity dependent cut-off, $z_{c}(L_{\rm X})$, after which the density drops as $(1+z)^{p2}$.  Unlike \citet{ueda2003} and others \citep{hasinger2005,silverman2008}, we do not fix any of these evolution parameters at a constant value.  While this complicates the procedure slightly, it allows more flexibility in fitting the data.

\subsection{Results for the Distant Sample}
\label{sec:results-fits}

In Table~\ref{tab:m1} we give the maximum likelihood fit parameters obtained using the ILDE model (which we only fit over $0<z<1.2$) for all spectroscopically identified sources and for broad-line AGNs.  We do the fits over the luminosity interval $10^{42}<L_{\rm X}<10^{46}$~ergs~s$^{-1}$.  The parameters for all spectral types are consistent with \citet{barger2005}, given the uncertainties.  While this is true of both the $z=0$ and evolution parameters, it is worth noting that the values of $pD=-0.94^{+0.74}_{-0.75}$ for the full sample and $pD=-0.81^{+0.63}_{-0.62}$ for the broad-lines do suggest some density evolution.  We plot the fits (all spectral types - \textit{blue solid}; broad-line AGNs - \textit{red dashed}) in Figures~\ref{fig:lfA}b$-$d and the $z=0$ extrapolations in Figure~\ref{fig:lfA}a (note that these fits do not include the local data shown in (a)).  We also include the $z=0$ extrapolation of the full sample fit in each of Figures~\ref{fig:lfA}b$-$d (\textit{black dotted}).  Visually, the curves provide a plausible, albeit simplified, match to the binned data, including the local luminosity function.  While the K-S tests indicate that the ILDE model may not adequately fit the unbinned data for all spectral types, the fit to the unbinned data for broad-line AGNs is much better, with 1-D K-S tests on the luminosity and redshift distributions at $91\%$ and $81\%$, respectively, and a 2-D test at $57\%$.  

In Table~\ref{tab:m2} we give the maximum likelihood fit parameters obtained using the LDDE model for all spectroscopically identified sources and for broad-line AGNs.  In each of the two cases we have fitted over three redshift intervals ($0<z<1.2$, $0<z<3$, and $0<z<5$).  We give the results for each interval in the table.  While the parameters of the $0<z<3$ and $0<z<5$ fits to all spectral types are generally consistent with each other, the parameters of the $0<z<1.2$ fit to all spectral types are typically quite different, reflecting the strong negative density evolution that occurs at higher redshifts.  In fact, it is the value of $p2$, which characterizes the high-redshift density evolution, that most distinguishes each of the three fits, showing a continual decrease as higher redshift data are included.  

Of our two high-redshift fits, the $0<z<3$ fit to all spectral types shows the most consistency with other HXLF studies that use an LDDE model \citep{ueda2003,lafranca2005,ebrero2008,silverman2008}.  This redshift interval is comparable to those studies, as \citet{ueda2003}, \citet{ebrero2008}, and \citet{silverman2008} have fits out to $z=3$, while \citet{lafranca2005} have fits out to $z=4$.  We show the most consistency with \citet{silverman2008}, significantly differing only in the overall normalization $A$.  Our value of $\log A = -6.18 \pm 0.04$ is noticeably lower than their $\log A = -6.08 \pm 0.02$.

It is important to note that, like \citet{silverman2008}, we do not correct for absorption in our HXLF model fits.  This is in contrast to the works of \citet{ueda2003}, \citet{lafranca2005}, and \citet{ebrero2008}, all of whom fit for the effects of absorption separately and determine ``intrinsic'', unabsorbed luminosity functions.  In spite of this, we typically show good agreement with most of their best-fit parameters, given the uncertainties. However, there are some notable exceptions.  Similar to our disagreement with \citet{silverman2008}, our estimate of $A$ is lower than each of their values.  We find our value of $\log L_{*}=44.28^{+0.15}_{-0.28}$ (in units of ergs~s$^{-1}$) to be larger than that\footnote{We have converted all of the $2-10$ keV luminosity parameters of \citet{ueda2003}, \citet{lafranca2005}, and \citet{ebrero2008} to $2-8$ keV luminosities assuming $\Gamma = 1.8$.} found by \citet{ebrero2008}, $\log L_{*}=43.83^{+0.01}_{-0.02}$, by $\sim 2 \sigma$, though consistent with $\log L_{*}=44.17 \pm 0.18$ of Model 4 of \citet{lafranca2005} and nearly with that of \citet{ueda2003}, $\log L_{*}=43.86^{+0.21}_{-0.26}$, given the uncertainties on these values.  Also, our value of $\log L_{a}=44.68^{+0.11}_{-0.12}$ is nearly consistent with the fixed value $\log L_{a}=44.52$ used by \citet{ueda2003} and \citet{ebrero2008}, though \citet{lafranca2005} fit this parameter freely and find a larger value of $\log L_{a}=45.68^{+0.58}_{-0.63}$.  

\begin{figure}
\epsscale{0.8}
\plotone{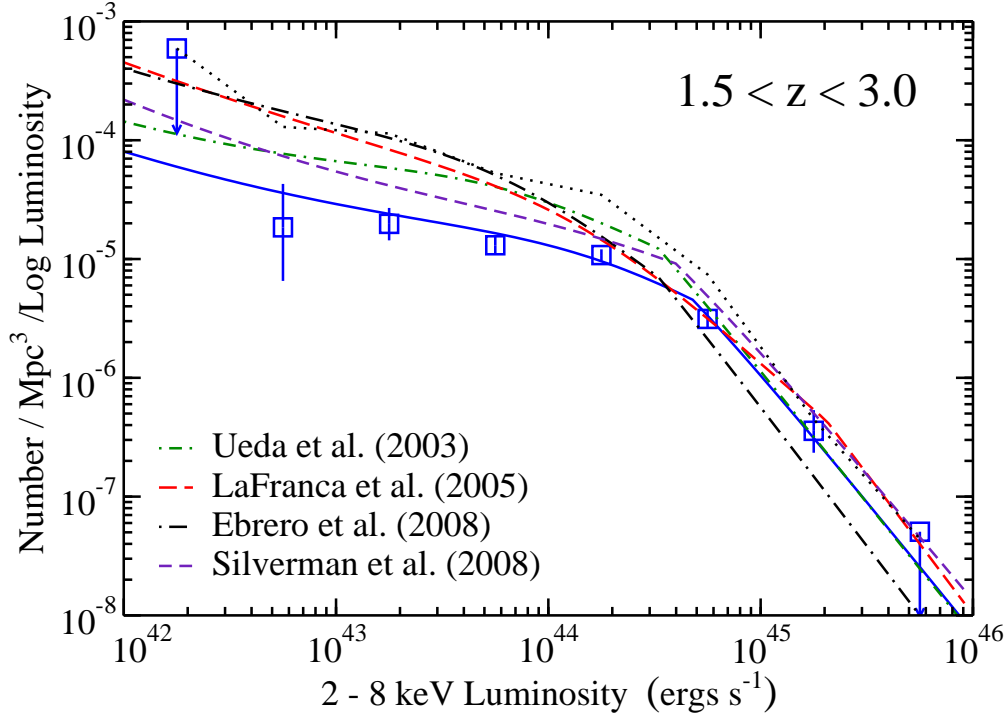}
\caption{The squares give the binned rest-frame $2-8$ keV luminosity function in the redshift region $1.5<z<3.0$.  Error bars indicate $1 \sigma$ Poissonian uncertainties on the number of objects in each bin.  The blue solid line is the best-fit LDDE model of our distant sample over the redshift region $0<z<3$, plotted at the geometric mean of the redshift bin, and the black dotted line gives the maximal binned HXLF for this redshift region (found by assigning the central redshift of the bin to all unidentified sources in the sample).  The other curves give the best-fit LDDE models of other studies, corrected from $2-10$ keV to $2-8$ keV luminosities assuming $\Gamma = 1.8$ and plotted at the geometric mean of the redshift bin (\citealp{ueda2003} - \textit{green short dashed-dotted}; \citealp{lafranca2005} - \textit{red long dashed};  \citealp{ebrero2008} - \textit{black long dashed-dotted}; \citealp{silverman2008} - \textit{purple short dashed}). }
\label{fig:faint}
\end{figure}

Of interest is how any disagreements in best-fit parameters are manifest in the $z=1.5-3.0$ region of the HXLF, particularly at the faint end where it is difficult to make accurate redshift measurements.  In Figure~\ref{fig:faint} we show our binned HXLF in this redshift bin (\textit{blue squares}), along with our best-fit $0<z<3$ LDDE model (\textit{blue solid curve}).  This is plotted at the geometric mean of the redshift bin, along with the best-fit LDDE models of \noindent \citeauthor{ueda2003}~(\citeyear{ueda2003}; \textit{green short dashed-dotted}), \citeauthor{lafranca2005}~(\citeyear{lafranca2005}; \textit{red long dashed}), \citeauthor{ebrero2008}~(\citeyear{ebrero2008}; \textit{black long dashed-dotted}), and \citeauthor{silverman2008}~(\citeyear{silverman2008}; \textit{purple short dashed}).  As described above, the shape of our HXLF agrees very well with that of \citet{silverman2008} but differs in the overall normalization.  Our fit also matches very well with that of \citet{ueda2003} at the bright end of the luminosity function, but their fit lies above ours below the luminosity break.  This discrepancy at the faint end is even more apparent in the curves of \citet{lafranca2005} and \citet{ebrero2008}.  It is also clear from Figure~\ref{fig:faint} that the LDDE fits of each of these studies lies above our binned determinations of the HXLF at the faint end, while showing some agreement at the bright end.  The incompleteness of our sample at faint fluxes is likely a large factor in these differences, as discussed in \S\ref{sec:results-bin}.  Our closer agreement with \citet{silverman2008} suggests that some discrepancy with the other studies may also come from trying to directly compare our HXLFs with their absorption corrected, intrinsic HXLFs.  This can also be seen by comparing the curves to the maximal binned HXLF in the $z=1.5-3.0$ bin, found by assigning all the spectroscopically observed but unidentified sources with valid flux measurements a redshift at the center of the bin.  This is given by the black dotted curve in Figure \ref{fig:faint}.  While the HXLFs of \citet{lafranca2005} and \citet{ebrero2008} are similar to the maximal HXLF at low luminosities, they do not match the overall shape for all luminosities, and the HXLFs of \citet{silverman2008} and \citet{ueda2003} appear to agree with the maximal curve only at higher luminosities.

%
% Figure 10 (faint15.eps)
%
\clearpage

In addition to their fit over $0.2<z<3.0$, \citet{silverman2008} find best-fit LDDE parameters over $0.2<z<5.5$.  Their sample contains $31$ AGNs with $z>3$, while our soft band sample has $41$ AGNs with $z>3$ (2 of which lie above $z=5$).  \citet{silverman2008} choose to fix all but 4 of their parameters (with the others set to the values found by their $0.2<z<3.0$ fit), while all of the parameters in our $0<z<5$ fits are allowed to vary freely.  This makes a direct comparison somewhat difficult.  However, they find a value of $A$ which is in very good agreement with our own.  They also find that a steeper value of $p2$ is necessary to match the high-redshift data, with their estimate of $p2=-3.27^{+0.31}_{-0.34}$ showing consistency with our $p2=-2.83^{+0.23}_{-0.24}$ within the uncertainties.  The only major disagreement between our fits is in the value of $\alpha$.  Their $\alpha=0.333 \pm 0.013$ is much larger than our $\alpha=0.208^{+0.014}_{-0.019}$.  This difference is likely caused by their fixing of $L_{a}$, a very complementary parameter.  Despite this, comparisons of the two LDDE models in the redshift bin $3<z<5$ show good agreement.

\begin{figure}
\epsscale{0.9}
\centering
\plottwo{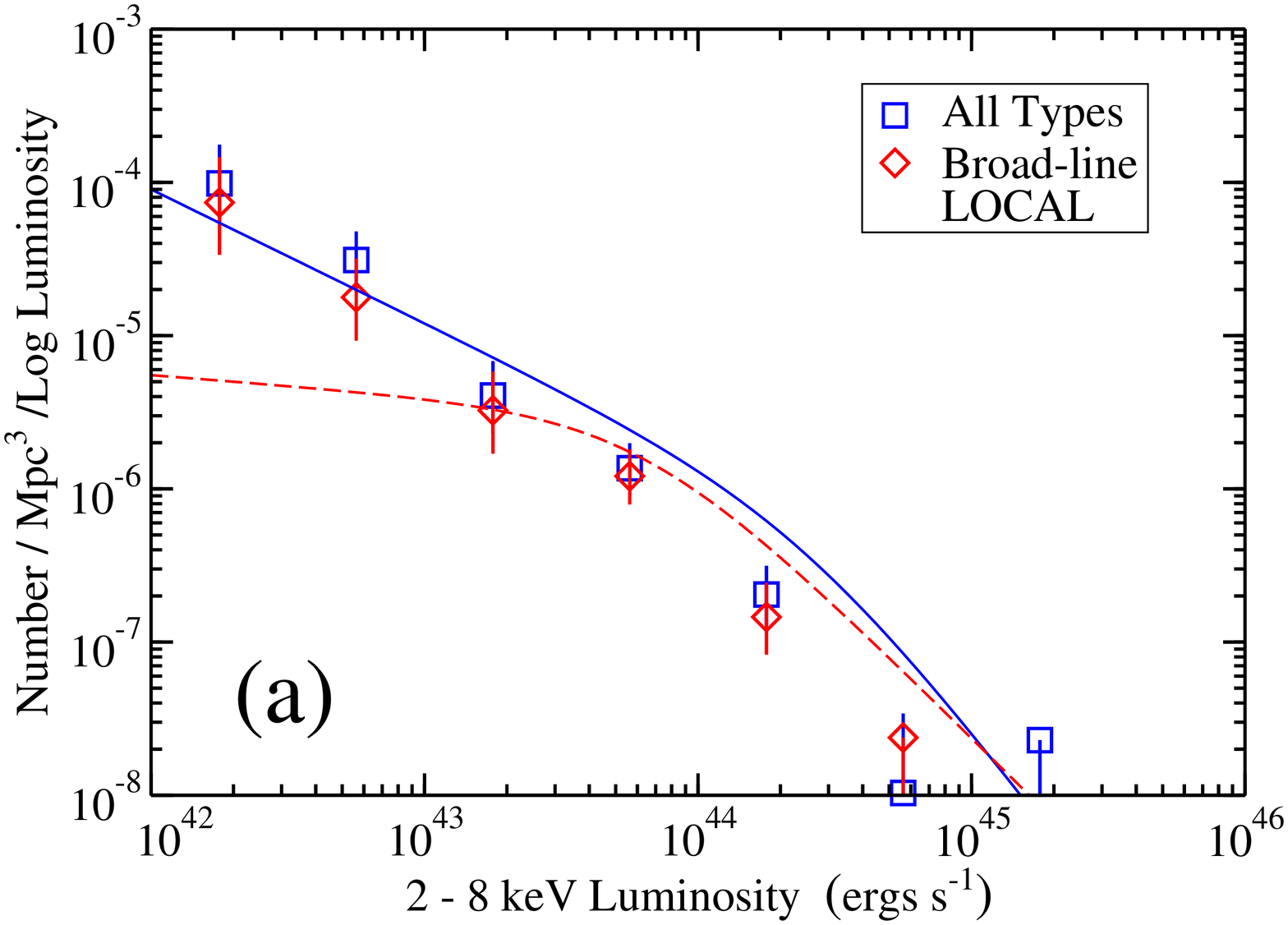}{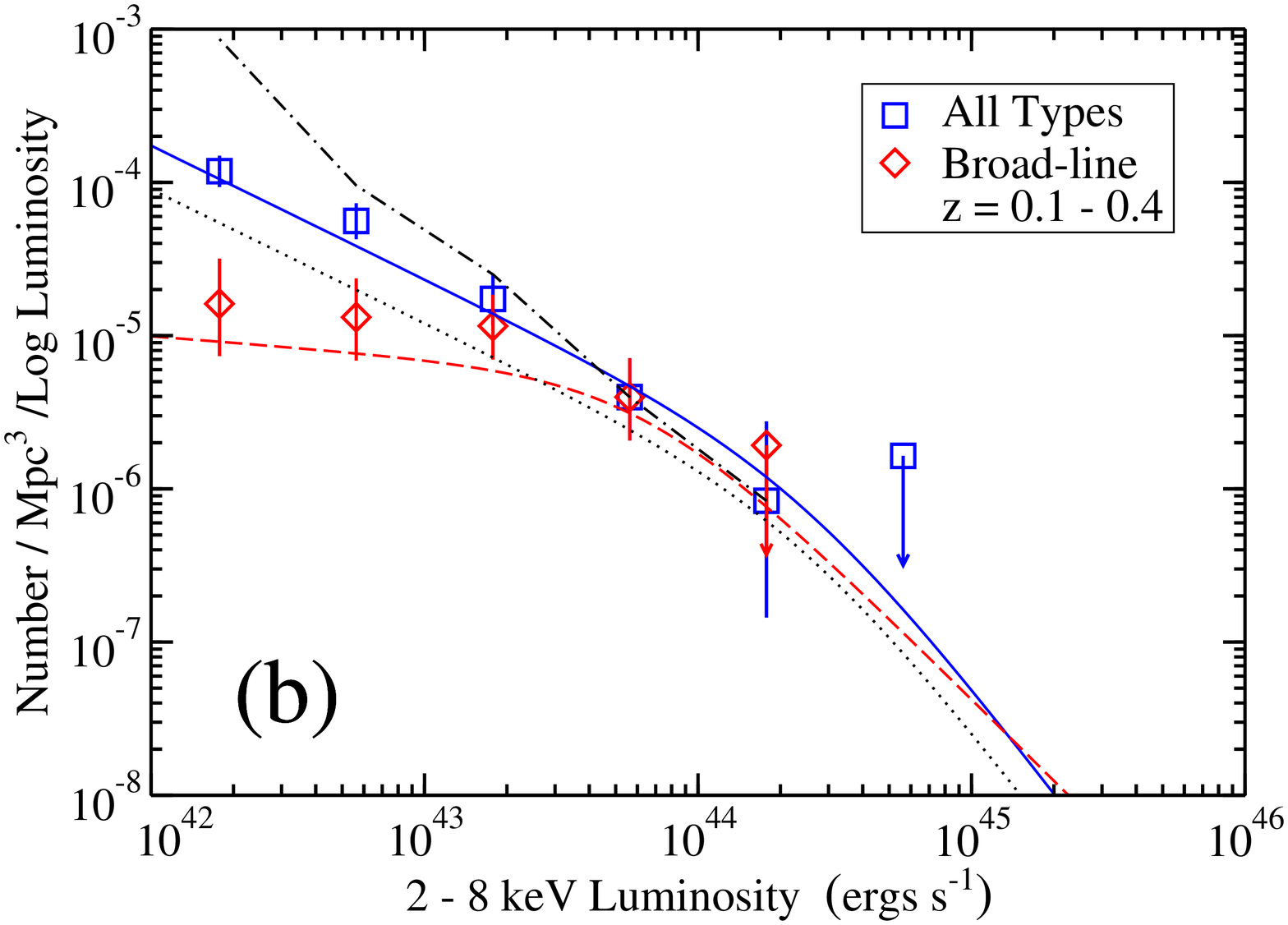}
\vskip 0.8cm
\plottwo{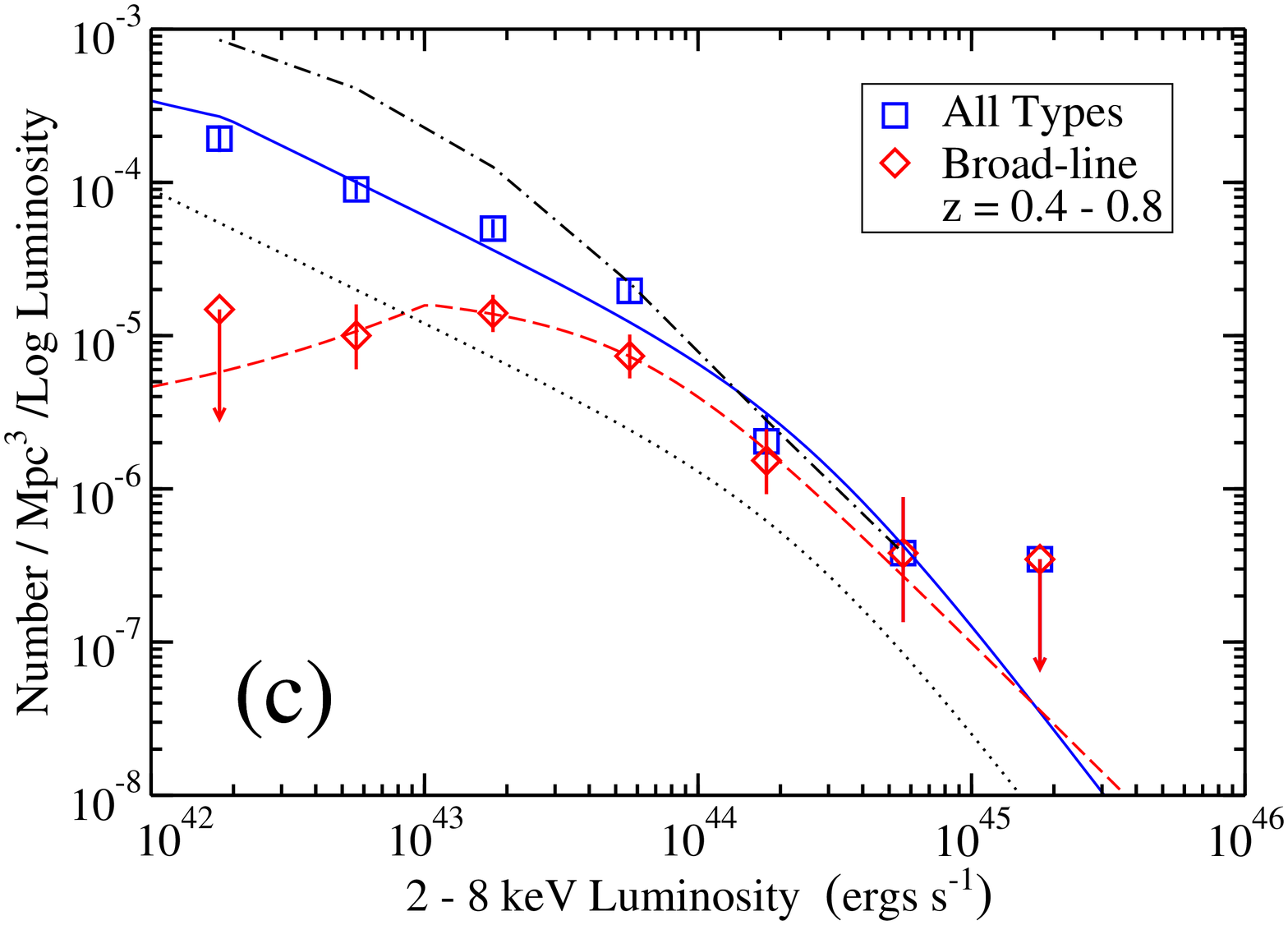}{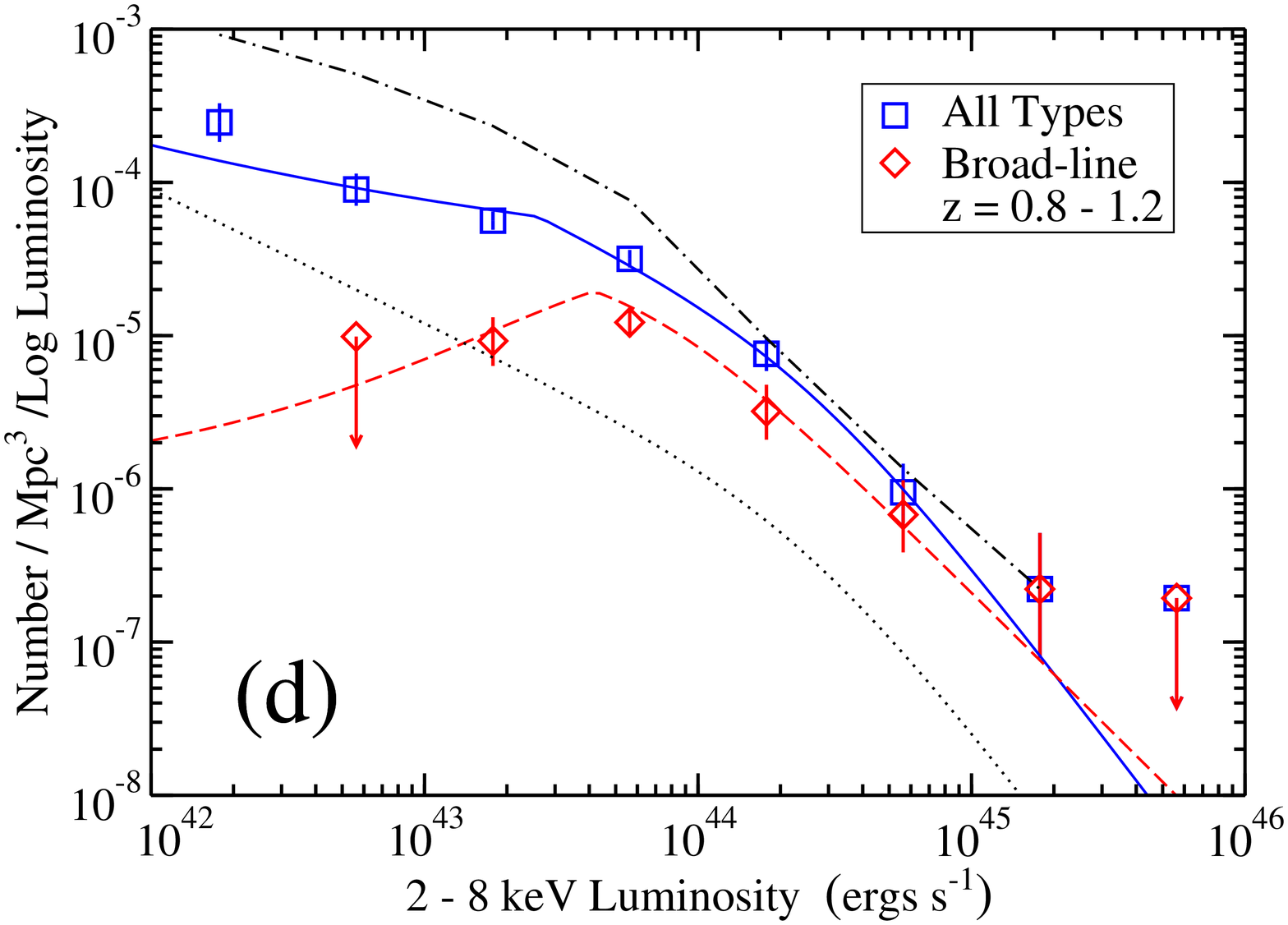}
\vskip 0.8cm
\plottwo{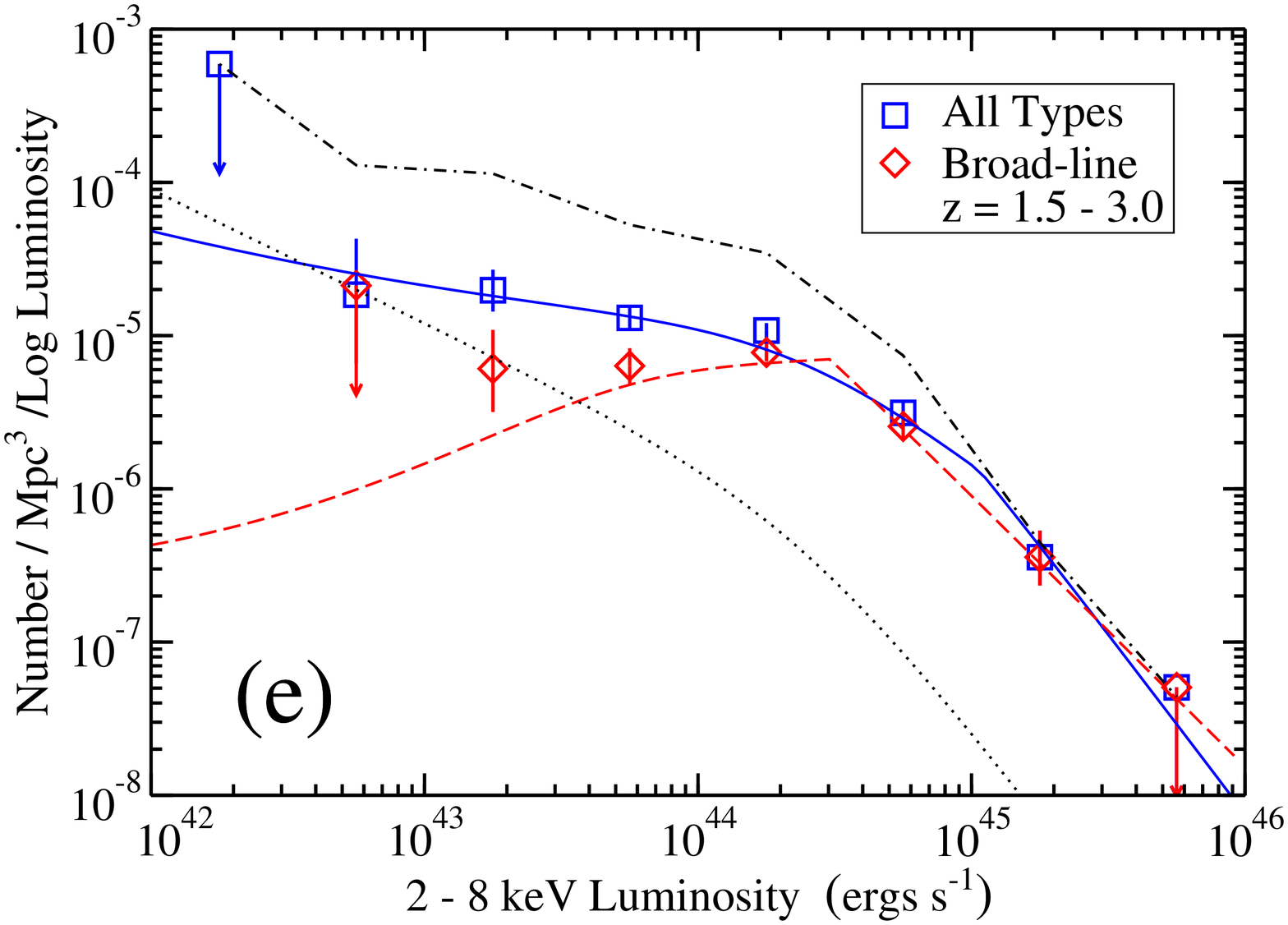}{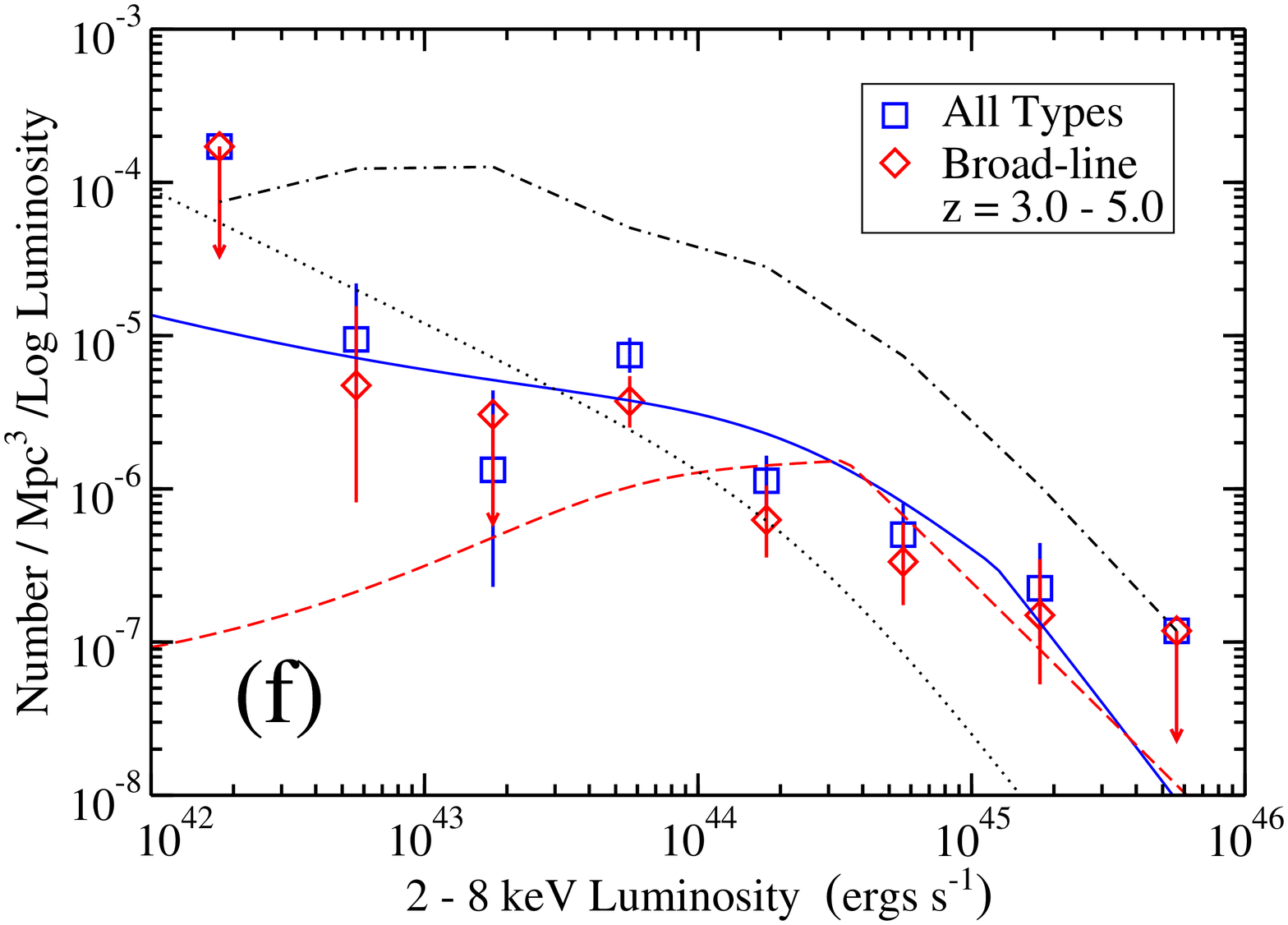}
\caption{
Binned rest-frame $2-8$ keV luminosity functions for our (a) $0<z<0.1$ local sample (see \S\ref{sec:localhxlf}) and for our distant sample in the low-redshift bins (b) $0.1 < z < 0.4$, (c) $0.4 < z < 0.8$, and (d) $0.8 < z < 1.2$ and in the high-redshift bins (e) $1.5 < z < 3.0$ and (f) $3.0<z<5.0$.  Blue squares (red diamonds) denote the HXLFs for all spectroscopically identified sources (broad-line AGNs) in each redshift bin.  Error bars indicate the 1~$\sigma$ Poissonian uncertainties based on the number of objects in each bin, while arrows denote $90\%$ (2.3~object) upper limits. The black dot-dashed curves in each panel (except a) give the maximal HXLFs found by assigning the central redshift of that redshift bin to all the spectroscopically observed but unidentified sources.  The blue solid and red dashed curves, respectively, show the maximum likelihood fits over $10^{42} < L_{\rm X} <10^{46}$~ergs~s$^{-1}$ and $0<z<5$ using the \textbf{LDDE model} (plotted at the geometric mean of each redshift bin) for the full and broad-line data of the distant sample.  The $z=0$ extrapolation of the fits is shown in (a), and the extrapolation of the full sample fit is also included in each higher redshift panel as a black dotted curve.}
\label{fig:lfD}
\end{figure}

To give a sense of the full redshift evolution of our fits, we plot our $0<z<5$ LDDE model fit in different redshift bins in Figure~\ref{fig:lfD} (all spectral types - \textit{blue solid}; broad-line AGNs - \textit{red dashed}).  We show the $z=0$ extrapolations in Figure~\ref{fig:lfD}a.  We also include the $z=0$ extrapolation of the full sample fit in each of Figures~\ref{fig:lfD}b$-$f (\textit{black dotted}).  (We note that our LDDE model fits over other redshift intervals show similar features.)  The fits to all spectral types show consistency with the binned luminosity functions over the full redshift interval fitted, but the $z=0$ extrapolations are in relatively poor agreement with the local luminosity function in Figure~\ref{fig:lfD}a.  The LDDE model fits in Figure~\ref{fig:lfD} reaffirm the notion that the broad-line AGN population density drops at low luminosities relative to that of the full population, particularly at higher redshifts.

% LDDE, high
%%
%% FIGURE 11  (old f10a-f.eps)
%%
\clearpage%%

%%
%% FIGURE 12  (old f11a-d.eps)
%%
\begin{figure} 
\epsscale{1.0}
\centering
\plottwo{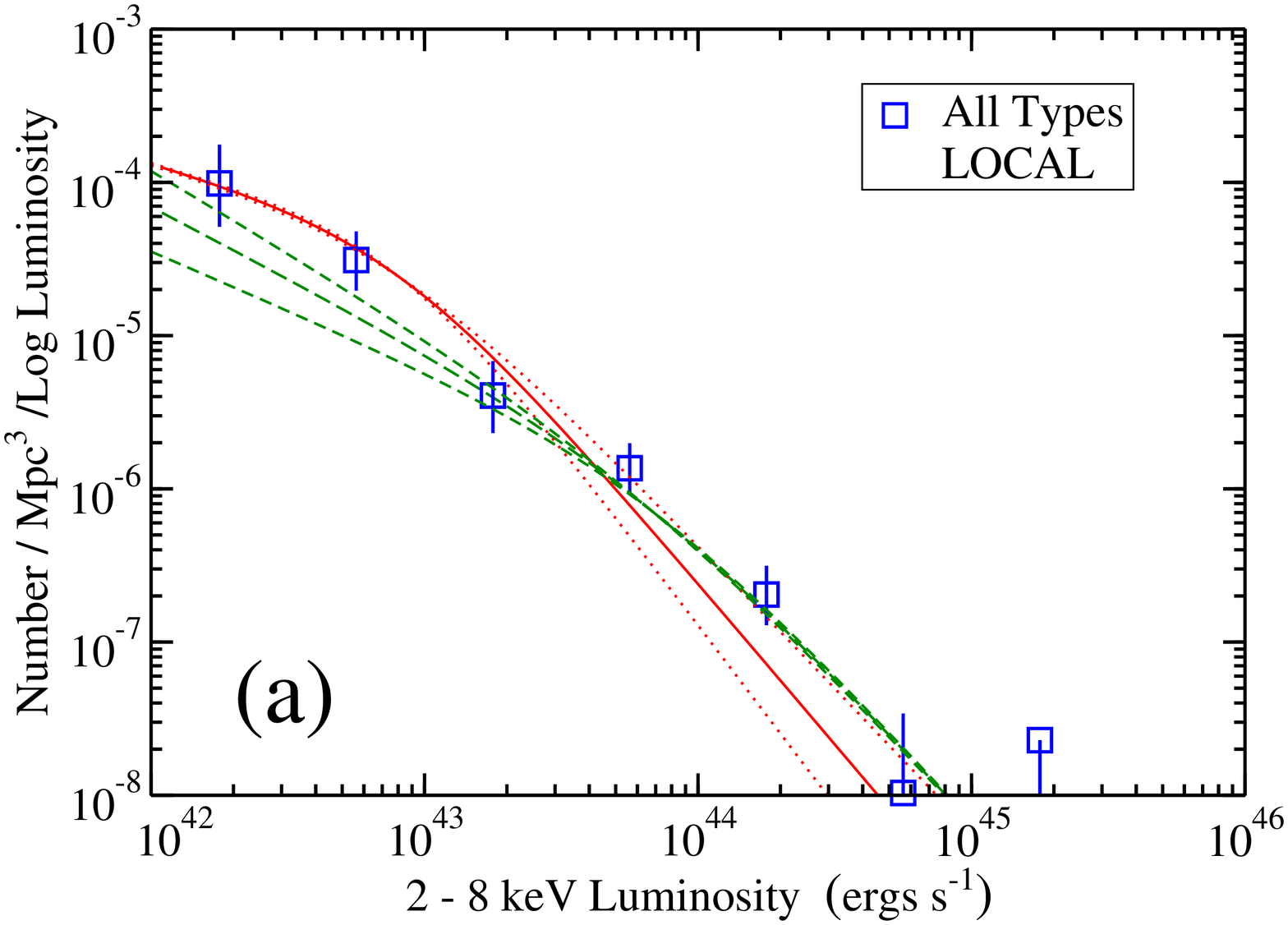}{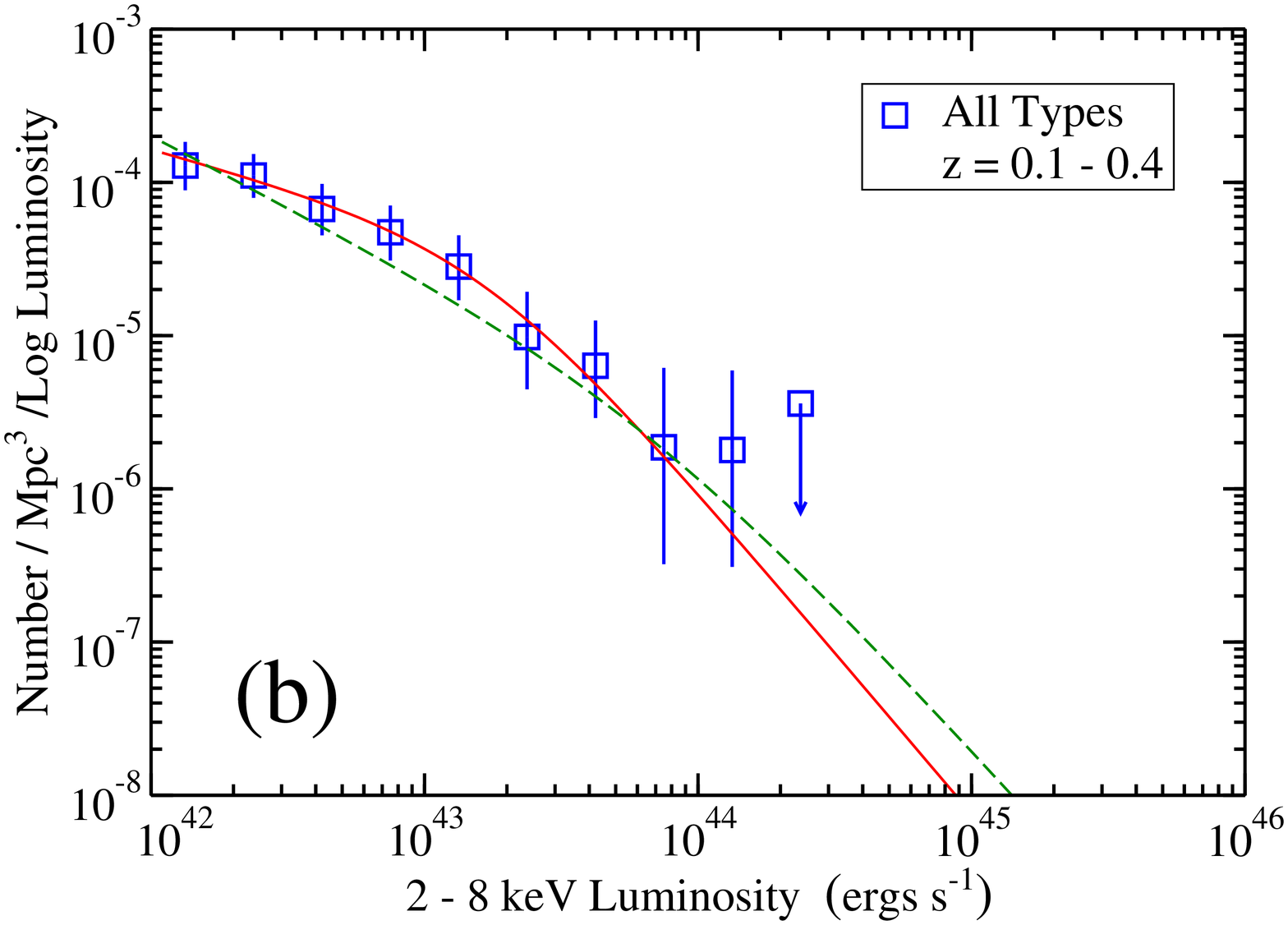}
\vskip 0.8cm
\plottwo{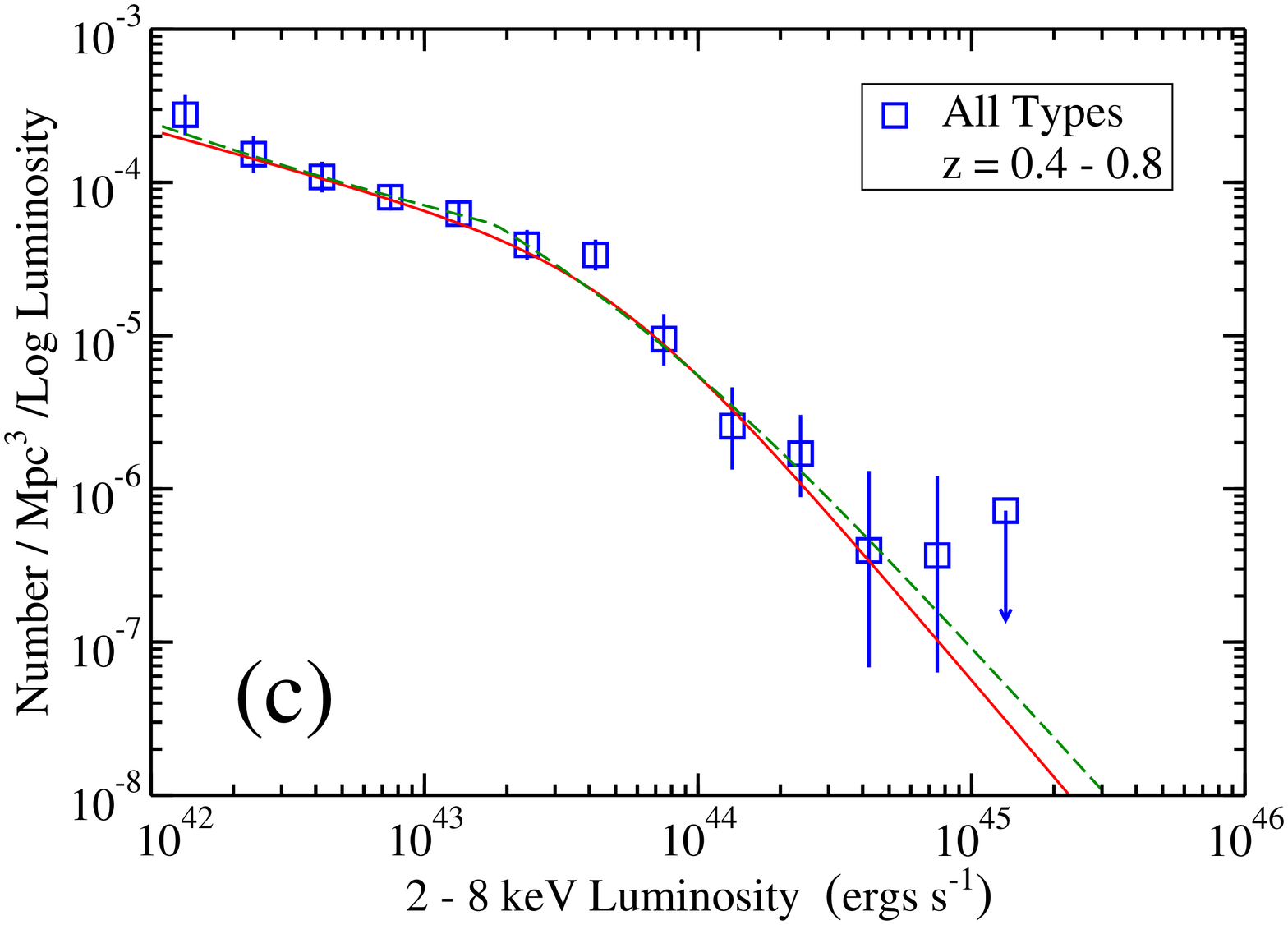}{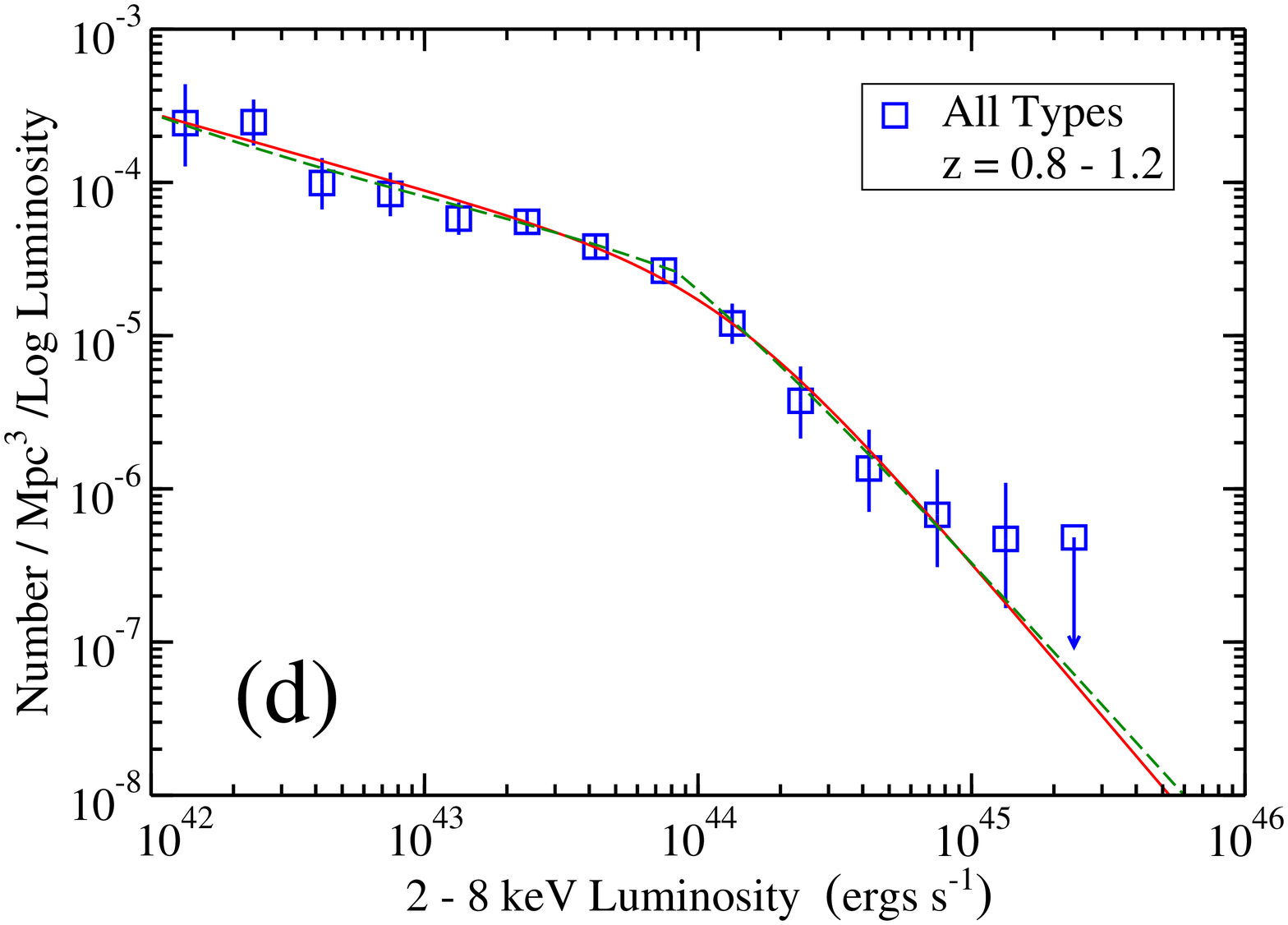}
\caption{Binned rest-frame $2-8$~keV luminosity functions for our (a) $0<z<0.1$ local sample (see \S\ref{sec:localhxlf}) and for our distant sample in the low-redshift bins (b) $0.1 < z < 0.4$, (c) $0.4 < z < 0.8$, and (d) $0.8 < z < 1.2$.  Blue squares denote the HXLFs for all spectroscopically identified sources in each redshift bin.  Error bars indicate the 1~$\sigma$ Poissonian uncertainties based on the number of objects in each bin, while arrows denote $90\%$ (2.3~object) upper limits. The red solid curve gives the fit to the distant sample over $0<z<1.2$ using the ILDE model, while the green long-dashed curve uses the LDDE model (the curves are shown at the geometric mean of each redshift bin).  The $z=0$ extrapolation of the fits is shown in (a).  Also in (a), the red dotted curves at the high luminosity end and the green short-dashed curves at the low luminosity end show the $1~\sigma$ spreads for the $z=0$ extrapolations using the ILDE model and the LDDE model, respectively.}
\label{fig:fine}
\end{figure}
\clearpage

While the fits to all spectral types are likely biased at high redshifts due to potentially large incompleteness, the broad-line AGNs are relatively free from such effects.  The LDDE model fits to the broad-line data over $0<z<3$ and $0<z<5$ show good agreement with the binned estimates, as seen in Figure \ref{fig:lfD} for the $0<z<5$ case, and the K-S tests suggest very good fits to the unbinned data.  The K-S tests are also favorable for the fit to the broad-line data over $0<z<1.2$.  However, the best-fit values of $z_{c}^{*}$, $L_{a}$, and $\alpha$ for the $0<z<1.2$ fit are quite different than those for the two high-redshift fits, and this results in curves which are rather cuspy and unlikely to be a good description of the actual population of low-redshift broad-line AGNs.  In general, we take the LDDE model fit over $0<z<3$ to be our best parameter set for broad-line AGNs in the distant sample.

Figures~\ref{fig:fine}b$-$d compare the fits over $0<z<1.2$ of the ILDE (\textit{red, solid curve}) and the LDDE (\textit{green, long-dashed curve}) models to all spectral types for a finer binned HXLF.  The models agree very well in the $z=0.8-1.2$ bin but begin to diverge in the lower redshift bins.  This can also be seen in Figure~\ref{fig:fine}a, where we compare the $z=0$ extrapolations of the fits (and their $1~\sigma$ spread) with our local HXLF.  Here it can also be seen that the extrapolation of the fit using the ILDE model matches the HXLF better at low luminosities, while the extrapolation of the fit using the LDDE model matches the HXLF better at high luminosities.

\subsection{Results for the Distant Plus Local Sample}
\label{sec:results-fits2}

We repeated our analysis on the combined distant plus local sample.  In Table \ref{tab:m1_bat} we give the maximum likelihood fit parameters obtained using the ILDE model (which we again only fit over $0<z<1.2$) for all spectroscopically identified sources and for broad-line AGNs.  We do the fits over the luminosity interval $10^{42}< L_{\rm X} < 10^{46}$~ergs~s$^{-1}$.  The fit to all spectral types shows virtually no deviation from the fit in \S\ref{sec:results-fits} other than a reduction in the uncertainties of some of the parameters. There is some variation in the parameters for the fit to the broad-line AGNs compared to the fit in \S\ref{sec:results-fits}, but, with the exception of a slight increase in $\gamma_{2}$, these are not statistically significant.

In Table~\ref{tab:m2_bat} we give the maximum likelihood fit parameters obtained using the LDDE model for all spectroscopically identified sources and for broad-line AGNs.  In each of the two cases we have fitted over three redshift intervals ($0<z<1.2$, $0<z<3$, and $0<z<5$).  We give the results for each interval in Table~\ref{tab:m2_bat}.  For the fits of all spectral types, the parameters in Table~\ref{tab:m2_bat} are generally consistent with the parameters in Table~\ref{tab:m2} but suggest a lower value of $\log L_{*}$ and larger $\log A$ and $p1$.  The changes in the former two parameters are the result of bringing the HXLF into better consistency with the local data, while the latter corrects for the change this would bring about at higher redshifts.

There is also a general trend among the Table~\ref{tab:m2_bat} fits to broad-line AGNs for larger values of the faint and bright end slopes (governed by $\gamma_{1}$ and $\gamma_{2}$).  This would be a mild improvement over the Table~\ref{tab:m2} fits as it would bring the bright ends of the fits to the broad-line AGNs into greater consistency with the corresponding fits to the full sample.  The faint ends of the fits to the broad-line AGNs, though potentially steeper, still lie well below the full population.  The $0<z<1.2$ fit, already problematic when fitting the distant sample, is further complicated by the addition of the local data.  The $z=0$ parameters show a strong break with those of the high-redshift fits and the K-S tests indicate an unsatisfactory fit.  Fortunately, the $0<z<3$ and $0<z<5$ fits provide a better description, with ($\log L_{\rm X}$, $z$, 2-D) K-S probabilities in the $0<z<1.2$ region of $(0.97,0.59,0.54)$ and $(0.99,0.56,0.49)$, respectively.

\begin{figure}
\epsscale{0.8}
\centering
\plotone{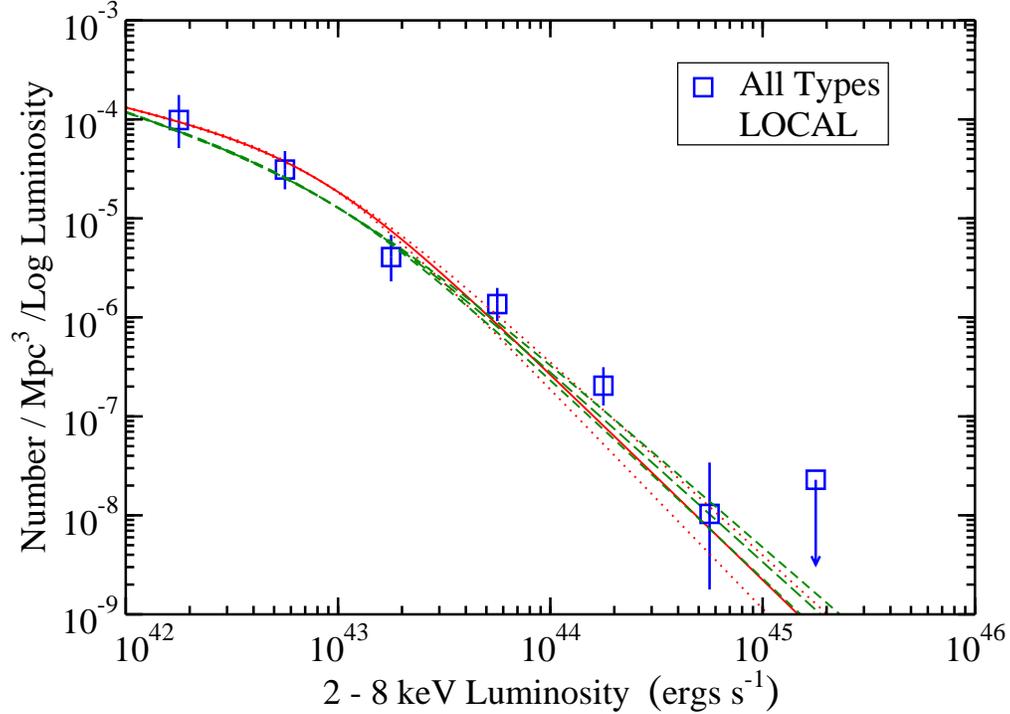}
\caption{Binned rest-frame $2-8$~keV luminosity function for our $0<z<0.1$ local sample (see \S\ref{sec:localhxlf}) (\textit{blue squares}).  Error bars indicate the 1~$\sigma$ Poissonian uncertainties based on the number of objects in each bin, while arrows denote $90\%$ (2.3~object) upper limits.  The red solid curve gives the $z=0$ extrapolation of the fit to the distant plus local sample over $0<z<1.2$ using the ILDE model, while the green long-dashed curve uses the LDDE model.  The red dotted curves and the green short-dashed curves show at the high luminosity end the $1~\sigma$ spreads for the $z=0$ extrapolations using the ILDE model and the LDDE model, respectively.}
\label{fig:allfine}
\end{figure}

In Figure~\ref{fig:allfine} we compare the fits over $0<z<1.2$ for the distant plus local sample using the ILDE (\textit{red solid curve}) and LDDE (\textit{green long-dashed curve}) models with the binned local HXLF.  Each model now agrees well with the local data at the faint end.  Additionally, we show the 1 $\sigma$ spreads on the bright end slope to demonstrate the consistency with the binned data in that region as well.  At higher redshifts, the LDDE model fits are very similar to those shown in Figure~\ref{fig:lfD}.
 
Combining the local sample with the distant sample generally improves the quality of the LDDE model fits over all redshift ranges.  We therefore recommend using the LDDE fits of the combined sample over $0<z<3$ for all the spectroscopically identified sources and for the broad-line AGNs as our best overall fits.  The LDDE fits over $0<z<5$ are also quite reasonable and could be useful when a description to very large redshifts is needed.  However, it must be noted that when dealing with the sample of all identified sources, the fits may be biased by incompleteness, with the greatest effects occurring for $z \gtrsim 1.5$.  For this reason, if one is only interested in redshifts $z \lesssim 1.2$, then the LDDE fit of the combined sample over $0<z<1.2$ for all identified sources (but not for broad-line AGNs) is likely to provide a better description of the data.

%%
%% FIGURE 13  (old f12.eps)
%%
\clearpage%%
\clearpage

\subsection{Space Density Evolution}

In addition to the luminosity function, the comoving space density of AGNs as a function of redshift can be useful in understanding the evolution of both the full and broad-line AGN populations.  These can be derived from an unbinned HXLF (which is simply the comoving space density per unit logarithmic luminosity) by integrating over the luminosity region of interest at a particular redshift.  In principle, a binned estimate can also be found by first computing a binned HXLF and multiplying by the size of the luminosity bin.  We will not follow this approach here, as our method of estimating the binned HXLF as described in \S\ref{sec:bin} assumes that it varies little over the bin, and we will be choosing luminosity bins large enough such that this assumption breaks down.

We will therefore estimate the comoving space density according to the traditional $1/V_{\rm max}$ method \citep{schmidt1968,felten1976,avni1980}.  At a redshift $z$ in a bin $z_{1} < z < z_{2}$ and in the luminosity range $\log L_{1} < \log L_{\rm X} < \log L_{2}$ this method gives

\be
\rho(z) \approx \sum_{i} \frac{1}{V_{\rm max, i}}
\ee

\noi where the sum is over all objects in this region of $(\log L_{X},z)$ space and

\be
V_{\rm max,i} = \int_{z_{1}}^{z_{2}} \Omega(\log L_{\rm X, i},z') \der{z}{V}(z')
\, \ud{z'}
\ee

\noi for a source with luminosity $\log L_{\rm X, i}$.  The factor $V_{\rm max,i}$ is the maximum accessible volume of each source as dictated by its luminosity and the survey details incorporated in $\Omega(\log L_{\rm X},z)$ (see \S\ref{sec:omega} for details).  The estimate of $\rho(z)$ is then just the sum of the individual contributions to the total space density.

\begin{figure}
\epsscale{0.6}
\plotone{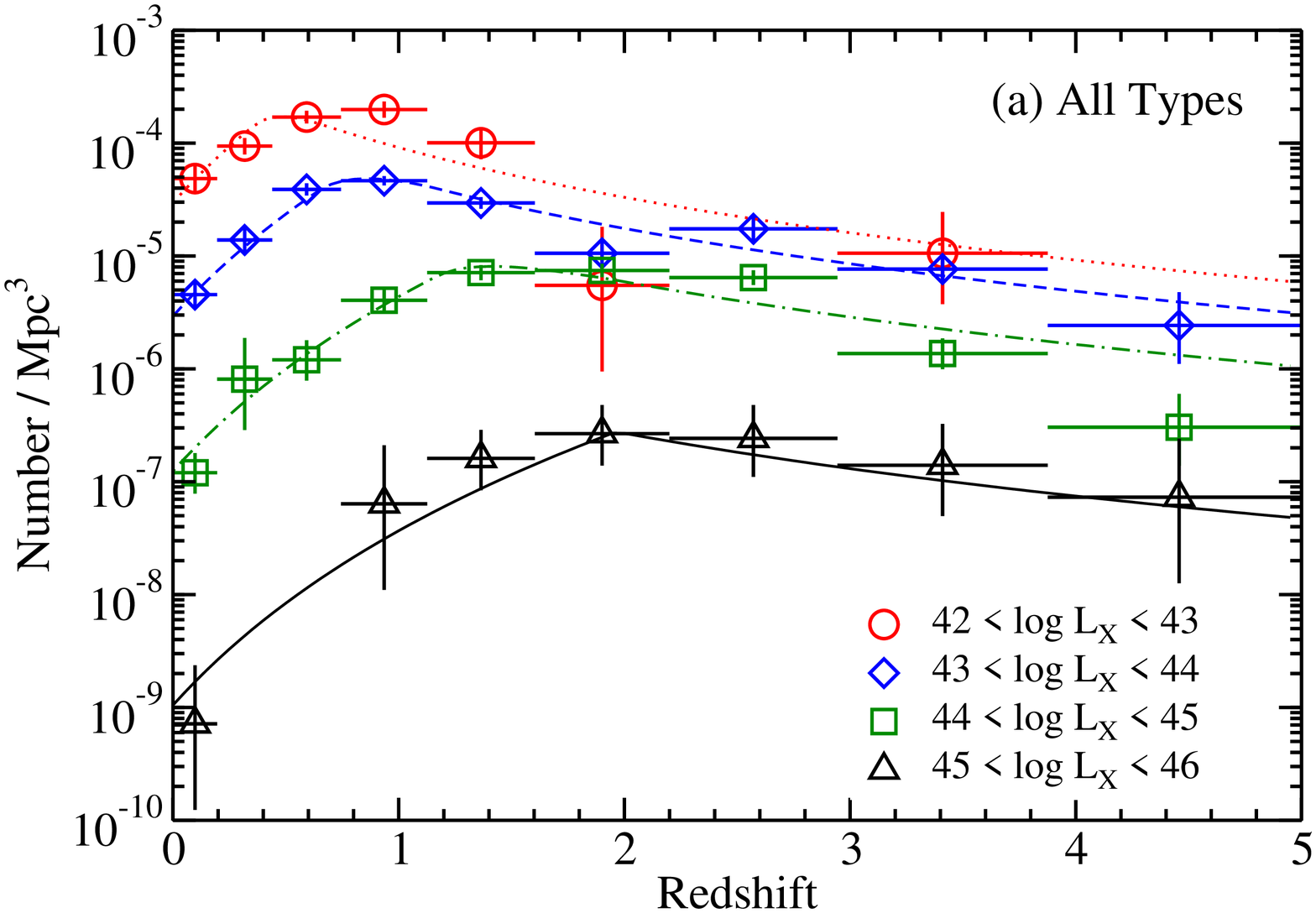} 
\vskip 1.2cm
\plotone{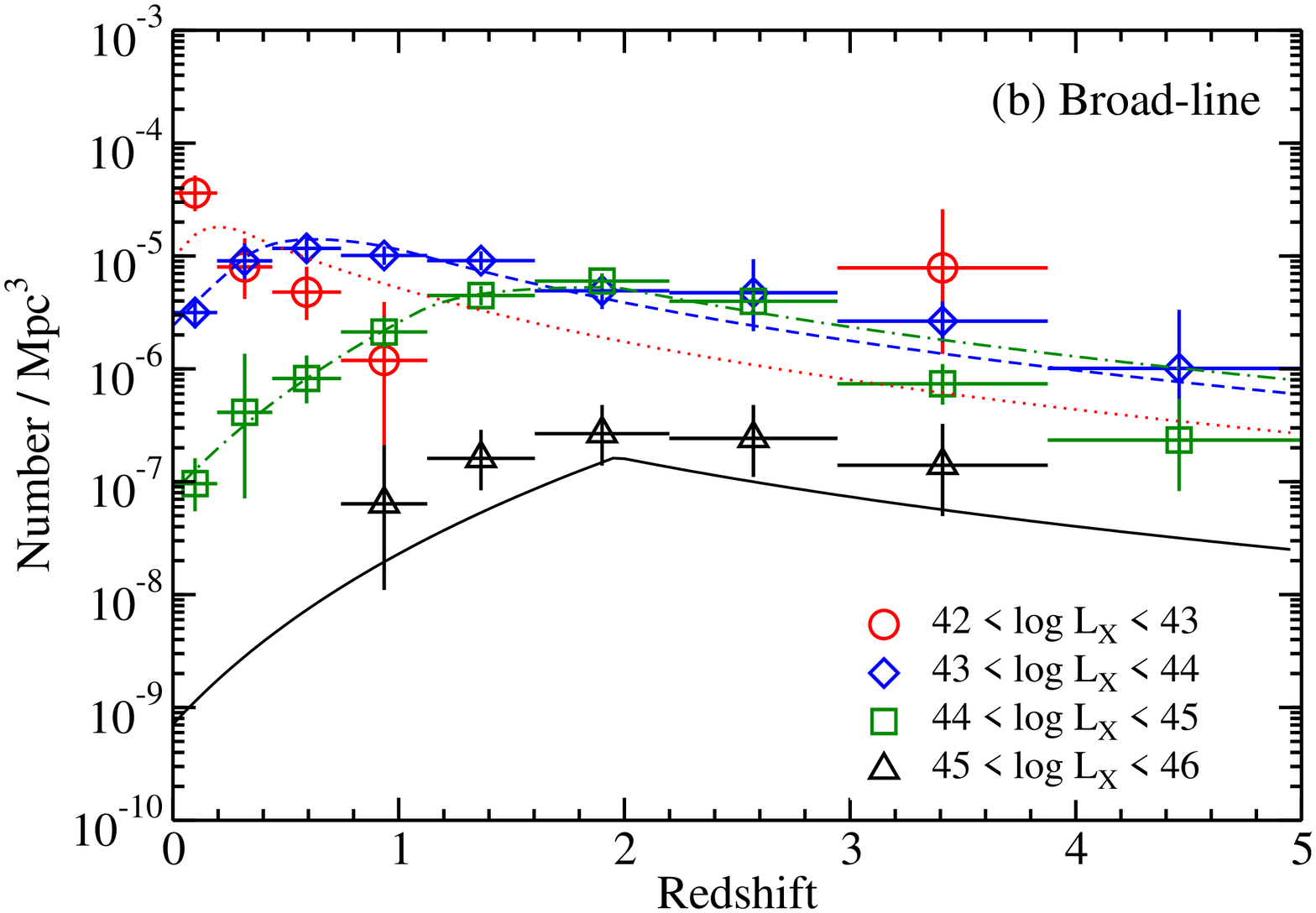} 
\caption{Comoving space density in the given luminosity intervals for (a) all spectral types and (b) broad-line AGN only.  The error bars indicate the $1 \sigma$ Poissonian uncertainties based on the number of objects in each bin.  The curves give the LDDE model fit of the HXLF (using the distant plus local sample over $z=0-5$ for the appropriate population) integrated over the intervals $\log L_{\rm X}=42-43$ (red dotted), $\log L_{\rm X}=43-44$ (blue dashed), $\log L_{\rm X}=44-45$ (green dashed-dotted), and $\log L_{\rm X}=45-46$ (solid black).}
\label{fig:evo}
\end{figure}

\begin{figure}
\epsscale{0.8}
\plotone{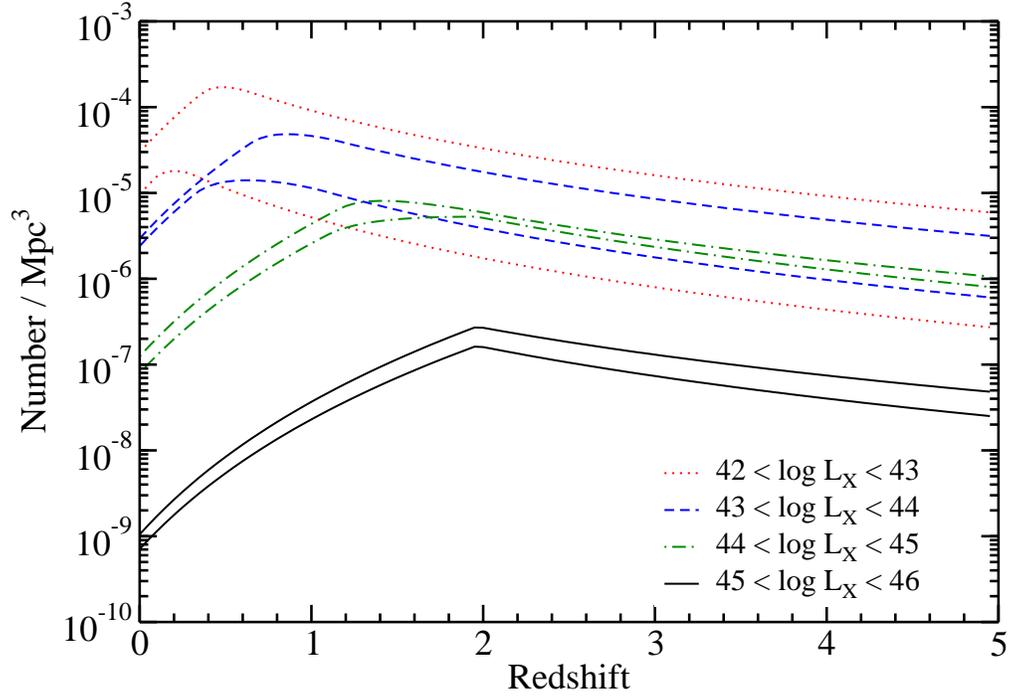} 
\caption{The upper (lower) curve of each pair gives the comoving space density for the full (broad-line) AGN population for each luminosity interval.  They are determined by integrating the LDDE model fit of the HXLF (using the distant plus local sample over $z=0-5$) over the intervals $\log L_{\rm X}=42-43$ (red dotted), $\log L_{\rm X}=43-44$ (blue dashed), $\log L_{\rm X}=44-45$ (green dashed-dotted), and $\log L_{\rm X}=45-46$ (solid black).}
\label{fig:evocomp}
\end{figure}

In Figures \ref{fig:evo}a and \ref{fig:evo}b we plot our binned estimates for the comoving space density for the full and broad-line populations, respectively, in four luminosity regions.  The errors are estimated as the $1 \sigma$ Poissonian uncertainties based on the number of objects in each bin.  Also plotted are the curves derived from integrating the unbinned HXLF over the corresponding luminosity intervals.  We use the LDDE model with parameters determined by the fit of the local plus distant sample over $z=0-5$, as given in Table \ref{tab:m2_bat}.  For both populations it is clear that the peak density of higher luminosity sources occurs at higher redshifts, and therefore earlier, than that of the lower luminosity sources.  This supports the idea of ``cosmic downsizing'' \citep{cowie1996} for AGNs \citep{cowie2003,fiore2003,steffen2003,ueda2003,barger2005,hasinger2005,lafranca2005,silverman2008}.  A direct comparison of the full and broad-line best-fit curves displayed separately in Figures \ref{fig:evo}a and \ref{fig:evo}b is given in Figure \ref{fig:evocomp}.  It is clear that while the broad-line population makes up a significant portion of the full population at higher luminosities for all redshifts, it is only below the redshifts at which the density peaks that faint broad-line sources make a sizable contribution.  This is similar to what is seen in the redshift evolution of the luminosity functions.

A close look at the binned data for the lowest luminosity region in Figure~\ref{fig:evo}b also reveals an interesting trend.  Unlike every other luminosity class for both the full and broad-line AGN samples, this class of broad-line AGNs is only now reaching its peak density\footnote{It should be noted that there is some disagreement between the binned data and the model fit in this luminosity region.  The discrepancy disappears, however, if the best-fit parameter values for the $z=0-1.2$ fit over the distant plus local sample are used, rather than the $z=0-5$ fit shown, as the $z=0-1.2$ fit is the most sensitive to the low-redshift data.}.  This is interesting to note in the context of the binned HXLF in Figure~\ref{fig:lfD}, in which it appears the low luminosity broad-line AGNs in the local sample have a noticeably larger density than in any of the higher redshift bins.  It is clear from Figure \ref{fig:evo}b, though, that the local data simply continue the upward trend in density and are not, in fact, atypical of the data as a whole.  As discussed in \S\ref{sec:results-bin}, we believe our broad-line identifications are robust and not significantly affected by host galaxy dilution, and that this is therefore an actual physical trend in the low luminosity broad-line AGN data.

%%
%% FIGURE 14 (evo1a/b.eps)
%%
\clearpage

%%
%% FIGURE 15 (evo1c.eps)
%%
\clearpage

\section{Summary}
\label{sec:summary}

In this paper we presented rest-frame hard ($2-8$~keV) X-ray luminosity functions constructed from a distant sample drawn from five highly spectroscopically complete surveys: CDF-N, CDF-S, CLASXS,  CLANS, and \emph{ASCA}.  Three of these surveys (CDF-N, CLASXS, and CLANS) compose our OPTX project \citep{trouille2008}.  The addition of the CLANS field substantially increases the sample over the one used by \citet{barger2005}. Following the general method of \citet{page2000}, we calculated binned HXLFs out to $z \approx 5$  for all the spectroscopically identified sources and for the broad-line AGNs.  We found that incompleteness is still a major source of uncertainty for the full (but not for the broad-line) HXLFs at $z \gtrsim 1.5$, potentially generating flatter HXLFs for lower luminosities than in reality.

A primary goal of our paper was to fit these data with evolutionary models. However, because fits using only distant data may extrapolate poorly to $z=0$, we also constructed a local $2-8$~keV HXLF selected from the very hard ($14-195$~keV) SWIFT BAT all-sky X-ray survey of \citet{tueller2008} so we could see how the inclusion of local data changes the model fits.  We used the maximum likelihood method with the combined distant plus local sample, as well as with the distant sample alone, to estimate parameters for two different types of models, one which allows for separate luminosity and density evolution (the ILDE model) and the other which provides for a luminosity dependent density evolution (LDDE).  

We found that the best-fit ILDE parameters over $0<z<1.2$ for the distant sample alone suggest some density evolution, particularly for the broad-line AGN population.  The best-fit LDDE parameters over $0<z<1.2$ also give a reasonable fit, further supporting density evolution at these low redshifts.  With the exception of the normalization $A$, our LDDE fit parameters over $0<z<3$ show very good agreement with those of \citet{silverman2008}.  There is also fairly good agreement with the ``intrinsic'' HXLF functions of \citet{ueda2003}, \citet{lafranca2005}, and \citet{ebrero2008}, with the exception of some discrepancies in the values of $L_{*}$, $A$, and $L_{a}$.  This causes some disagreement at the faint end of the HXLF for $z>1.5$, which is likely due to potentially significant incompleteness effects caused by the difficulty in spectroscopically observing high-redshift, low luminosity sources.  Fortunately, as we believe our broad-line AGN sample is relatively free of these complications and therefore fairly complete, we have best-fit LDDE parameters which provide a very good match to the broad-line AGN data up to $z \sim 3$ and a reasonable fit out to $z \sim 5$, as well.

The fits using the ILDE and LDDE models for the distant plus local sample are typically quite similar to the fits without the local data when considering the uncertainty on the parameters, but in some cases they do suggest some slight modifications.  The fits using the LDDE model for the spectroscopically identified sample over $0<z<3$ and $0<z<5$ are consistent with slightly lower values of the luminosity break $L_{*}$ and with larger values of the normalization $A$ and low-redshift density evolution parameter $p1$.  This brings the $z=0$ extrapolations into better agreement with the local binned data while maintaining the congruity at high redshifts.  Also, in all cases the fits of the broad-line AGN data show a mild increase in the values of the faint and bright end slopes, $\gamma_{1}$ and $\gamma_{2}$, which brings the bright ends of these fits into greater consistency with the fits of the full data while retaining the discrepancy at the faint end.  We believe these to be improvements over the fits of the distant sample alone and therefore recommend using the LDDE parameters of the combined sample over $0<z<3$ for all the spectroscopically identified sources and for the broad-line AGNs as our best overall fits, while still noting the possible effects of incompleteness on the fits of the full sample.

When examining both the binned an unbinned HXLFs, constructed with and without the local data, we confirm that at all redshifts the population of broad-line AGNs follows a fundamentally different form than that of the full population.  While the broad-line AGNs account for a majority of the overall AGN space density at high luminosities, their density drops significantly for lower luminosities.

\acknowledgments
The authors would like to thank the referee for comments and suggestions which helped to improve this manuscript.  A.~J.~B. thanks the Aspen Center for Physics for hospitality during the completion of this work.  We gratefully acknowledge support from NSF grants AST 0239425 and AST 0708793 (A.~J.~B.), the University of Wisconsin Research Committee with funds granted by the Wisconsin Alumni Research Foundation, and the David and Lucile Packard Foundation (A.~J.~B.).  L.~T. was supported by a National Science Foundation Graduate Research Fellowship and a Wisconsin Space Grant Consortium Graduate Fellowship Award during portions of this work.

%%
%% TABLE 3
%%
\clearpage% ILDE, all/blagn
\begin{deluxetable}{|c|c|c|}
%\begin{deluxetable}{ccc}
\tablewidth{0pt}
\tablecaption{Maximum Likelihood Fit Parameters for All Spectroscopically Identified Sources and for Broad-Line AGNs Using a ILDE Model\label{tab:m1}}
\startdata
\hline \hline
\multicolumn{3}{|l|}{ILDE} \\
%\multicolumn{3}{c}{ILDE} \\
%ILDE & & \\
\hline \hline
 & All Identified & BLAGN \\
\hline
 & ($0<z<1.2$) & ($0<z<1.2$) \\
\hline
$\log L_{*}$\tnm{a} & $42.91^{+0.25}_{-0.25}$    & $42.63^{+0.22}_{-0.21}$
\\
$\log A$\tnm{b}     & $-4.319^{+0.047}_{-0.050}$ & $-4.340^{+0.091}_{-0.101}$
\\
$\gamma_{1}$        & $0.498^{+0.081}_{-0.087}$  & $-1.08^{+0.39}_{-0.47}$
\\
$\gamma_{2}$        & $2.10^{+0.25}_{-0.23}$     & $1.48^{+0.20}_{-0.17}$
\\ \hline
$pL$                & $4.04^{+0.85}_{-0.86}$     & $3.54^{+0.66}_{-0.65}$
\\
$pD$                & $-0.94^{+0.74}_{-0.75}$    & $-0.81^{+0.63}_{-0.62}$
\\ \hline
%K-S Probabilities &  &  \\
$\rm P_{KS}$ & & \\
$(L_{x},z,2D)$ & $(0.93,0.003,0.05)$ & $(0.91,0.81,0.57)$ \\
\hline
\enddata
\tnt{a}{In units of ergs s$^{-1}$}
\tnt{b}{In units of Mpc$^{-3}$}
\end{deluxetable}

%%
%% TABLE 4
%%
% LDDE, all identified
\begin{deluxetable}{|c|ccc|ccc|}
\rotate
\tablewidth{0pt}
\tablecaption{Maximum Likelihood Fit Parameters For All Spectroscopically Identified Sources and for Broad-Line AGNs Using a LDDE Model\label{tab:m2}}
\startdata
\hline \hline
\multicolumn{7}{|l|}{LDDE} \\
\hline \hline
& \multicolumn{3}{c|}{All Identified} & \multicolumn{3}{c|}{BLAGN} \\
\hline
               & $(0<z<1.2)$             & $(0<z<3)$
               & $(0<z<5)$
               & $(0<z<1.2)$             & $(0<z<3)$
               & $(0<z<5)$
\\
\hline
$\log L_{*}$\tnm{a} 
                    & $43.83^{+0.36}_{-0.50}$     & $44.28^{+0.15}_{-0.28}$
                    & $44.40^{+0.13}_{-0.14}$
               & $44.09^{+0.87}_{-1.34}$      & $43.91^{+0.20}_{-0.25}$
               & $43.81^{+0.22}_{-0.16}$
\\
$\log A$\tnm{b}     
                    & $-5.846^{+0.047}_{-0.050}$ & $-6.176^{+0.037}_{-0.039}$
                    & $-6.140^{+0.036}_{-0.038}$
               & $-5.922^{+0.091}_{-0.101}$   & $-5.618^{+0.059}_{-0.063}$
               & $-5.510^{+0.056}_{-0.060}$ 
\\
$\gamma_{1}$        
                    & $0.92^{+0.13}_{-0.16}$     & $0.947^{+0.066}_{-0.085}$
                    & $0.872^{+0.060}_{-0.058}$
               & $0.71^{+0.32}_{-0.62}$      & $0.25^{+0.16}_{-0.21}$
               & $0.14^{+0.190}_{-0.19}$     
\\
$\gamma_{2}$        
                    & $1.98^{+0.30}_{-0.25}$     & $2.21^{+0.21}_{-0.21}$
                    & $2.36^{+0.24}_{-0.20}$
               & $1.72^{+0.98}_{-0.50}$    & $1.84^{+0.18}_{-0.17}$
               & $1.77^{+0.17}_{-0.15}$ 
\\ \hline
$p1$                
                    & $5.84^{+1.70}_{-0.89}$     & $4.29^{+0.58}_{-0.37}$
                    & $3.61^{+0.49}_{-0.33}$  
               & $3.31^{+0.85}_{-0.87}$       & $3.16^{+0.50}_{-0.41}$
               & $3.20^{+0.40}_{-0.49}$  
\\
$p2$                
                    & $0.56^{+0.49}_{-0.52}$     & $-1.70^{+0.31}_{-0.34}$
                    & $-2.83^{+0.23}_{-0.24}$ 
               & $-4.9^{+2.0}_{-1.8}$         & $-3.15^{+1.41}_{-0.70}$
               & $-3.45^{+0.41}_{-0.41}$
\\
$z_{c}^{*}$         
                    & $0.947^{+0.055}_{-0.142}$  & $1.80^{+0.13}_{-0.15}$
                    & $2.18^{+0.55}_{-0.29}$  
               & $1.022^{+0.044}_{-0.170}$    & $2.42^{+0.21}_{-0.46}$
               & $2.24^{+0.12}_{-0.18}$
\\
$\log L_{a}$\tnm{a} 
                    & $43.921^{+0.073}_{-0.109}$ & $44.68^{+0.11}_{-0.12}$
                    & $45.09^{+0.49}_{-0.37}$
               & $43.51^{+0.38}_{-0.11}$   & $44.57^{+0.10}_{-0.14}$
               & $44.539^{+0.054}_{-0.102}$
\\
$\alpha$            
                    & $0.337^{+0.069}_{-0.087}$  & $0.269^{+0.031}_{-0.026}$
                    & $0.208^{+0.014}_{-0.019}$
               & $0.92^{+0.37}_{-0.22}$       & $0.422^{+0.088}_{-0.037}$
               & $0.390^{+0.040}_{-0.033}$ 
\\ \hline
%K-S Probabilities & & & & & &
$\rm P_{KS}$  & & & & & &
\\
$(L_{x},z,2D)$ & $(0.99,0.04,0.23)$         & $(0.93,0.04,0.08)$
               & $(0.98,0.10,0.18)$
               & $(0.91,0.87,0.95)$           & $(0.97,0.96,0.96)$
               & $(0.98,0.59,0.43)$
\\
\hline
\enddata
\tnt{a}{In units of ergs s$^{-1}$}
\tnt{b}{In units of Mpc$^{-3}$}
\end{deluxetable}

%%
%% TABLE 5
%%
% ILDE, all/blagn w/ BAT data 
\begin{deluxetable}{|c|c|c|}
\tablewidth{0pt}
\tablecaption{Maximum Likelihood Fit Parameters for All Spectroscopically Identified Sources and for Broad-Line AGNs Using a ILDE Model (Distant Plus Local Sample)\label{tab:m1_bat}}
\startdata 
\hline \hline
\multicolumn{3}{|l|}{ILDE} \\
\hline \hline
 & All Identified & BLAGN \\
\hline
 & ($0<z<1.2$) & ($0<z<1.2$) \\
\hline
$\log L_{*}$\tnm{a} & $42.92^{+0.12}_{-0.16}$    & $42.70^{+0.18}_{-0.14}$
\\
$\log A$\tnm{b}     & $-4.328^{+0.046}_{-0.048}$ & $-4.267^{+0.083}_{-0.091}$
\\
$\gamma_{1}$        & $0.500^{+0.075}_{-0.081}$  & $-0.67^{+0.19}_{-0.33}$
\\
$\gamma_{2}$        & $2.08^{+0.14}_{-0.12}$     & $1.90^{+0.14}_{-0.12}$
\\ \hline
$pL$                & $3.98^{+0.47}_{-0.46}$     & $4.04^{+0.43}_{-0.22}$
\\
$pD$                & $-0.91^{+0.49}_{-0.52}$    & $-1.35^{+0.37}_{-0.53}$
\\ \hline
%K-S Probabilities &  &  \\
$\rm P_{KS}$  & & \\
$(L_{x},z,2D)$ & $(0.77,0.005,0.09)$ & $(0.43,0.90,0.41)$ \\
\hline
\enddata
\tnt{a}{In units of ergs s$^{-1}$}
\tnt{b}{In units of Mpc$^{-3}$}
\end{deluxetable}

%%
%% TABLE 6
%%
% LDDE, all identified w/ BAT data
\begin{deluxetable}{|c|ccc|ccc|}
%\begin{deluxetable}{ccccccc}
\rotate
\tablewidth{0pt}
\tablecaption{Maximum Likelihood Fit Parameters For All Identified Spectral Types and for Broad-Line AGNs Using a LDDE Model (Distant Plus Local Sample)\label{tab:m2_bat}}
\startdata
\hline \hline
\multicolumn{7}{|l|}{LDDE} \\
%\multicolumn{7}{l}{LDDE} \\
\hline \hline
& \multicolumn{3}{c|}{All Identified} & \multicolumn{3}{c|}{BLAGN} \\
%& \multicolumn{3}{c}{All Identified} & \multicolumn{3}{c}{BLAGN} \\
\hline
               & $(0<z<1.2)$             & $(0<z<3)$
               & $(0<z<5)$
               & $(0<z<1.2)$             & $(0<z<3)$
               & $(0<z<5)$
\\
\hline
$\log L_{*}$\tnm{a} 
                    & $42.99^{+0.27}_{-0.16}$     & $43.99^{+0.22}_{-0.42}$
                    & $44.078^{+0.037}_{-0.097}$
               & $44.64^{+0.31}_{-0.21}$      & $43.64^{+0.18}_{-0.19}$
               & $43.54^{+ 0.24}_{-0.13}$
\\
$\log A$\tnm{b}     
                    & $-4.582^{+0.046}_{-0.048}$ & $-6.060^{+0.036}_{-0.038}$
                    & $-6.117^{+0.035}_{-0.037}$
               & $-7.189^{+0.083}_{-0.091}$   & $-5.545^{+0.056}_{-0.060}$
               & $-5.425^{+0.055}_{-0.058}$ 
\\
$\gamma_{1}$        
                    & $0.685^{+0.242}_{-0.048}$     & $1.004^{+0.078}_{-0.087}$
                    & $0.956^{+0.094}_{-0.055}$
               & $1.196^{+0.142}_{-0.064}$      & $0.45^{+0.18}_{-0.18}$
               & $0.39^{+0.18}_{-0.19}$  
\\
$\gamma_{2}$        
                    & $1.936^{+0.084}_{-0.074}$     & $2.24^{+0.16}_{-0.20}$
                    & $2.30^{+0.28}_{-0.12}$
               & $3.122^{+1.40}_{-0.55}$    & $2.15^{+0.14}_{-0.11}$
               & $2.09^{+0.14}_{-0.12}$
\\ \hline
$p1$                
                    & $6.42^{+0.49}_{-0.32}$     & $5.58^{+0.41}_{-0.52}$
                    & $5.13^{+0.35}_{-0.21}$
               & $4.80^{+0.51}_{-0.49}$       & $4.93^{+0.29}_{-0.31}$
               & $5.02^{+0.43}_{-0.31}$
\\
$p2$                
                    & $1.11^{+0.39}_{-0.34}$     & $-1.34^{+0.29}_{-0.24}$
                    & $-2.51^{+0.25}_{-0.16}$
               & $-4.6^{+1.1}_{-1.6}$         & $-2.16^{+0.43}_{-0.43}$
               & $-2.70^{+0.28}_{-0.47}$
\\
$z_{c}^{*}$         
                    & $0.953^{+0.049}_{-0.053}$  & $1.69^{+0.18}_{-0.15}$
                    & $1.964^{+0.198}_{-0.047}$
               & $0.900^{+0.150}_{-0.030}$    & $2.23^{+0.11}_{-0.17}$
               & $1.96^{+0.19}_{-0.15}$  
\\
$\log L_{a}$\tnm{a} 
                    & $43.887^{+0.040}_{-0.051}$ & $44.68^{+0.24}_{-0.12}$
                    & $44.889^{+0.056}_{-0.115}$
               & $43.434^{+0.126}_{-0.024}$   & $44.566^{+0.031}_{-0.147}$
               & $44.478^{+0.091}_{-0.056}$
\\
$\alpha$            
                    & $0.694^{+0.143}_{-0.062}$  & $0.303^{+0.049}_{-0.023}$
                    & $0.255^{+0.016}_{-0.022}$
               & $1.28^{+0.14}_{-0.21}$       & $0.553^{+0.050}_{-0.026}$
               & $0.535^{+0.023}_{-0.077}$
\\ \hline
%K-S Probabilities & & & & & & 
%\\
$\rm P_{KS}$  & & & & & & \\
$(L_{x},z,2D)$ & $(0.97,0.02,0.14)$         & $(0.91,0.07,0.15)$
               & $(0.97,0.13,0.43)$
               & $(0.69,0.77,0.31)$           & $(0.99,0.86,0.85)$
               & $(0.97,0.44,0.40)$
\\
\hline
\enddata
\tnt{a}{In units of ergs s$^{-1}$}
\tnt{b}{In units of Mpc$^{-3}$}
\end{deluxetable}

\clearpage

%\newpage

\end{document}